\documentclass{agujournal2019}
\usepackage{url} 
\usepackage{soul}
\usepackage{amssymb}
\usepackage{pifont}
\usepackage{subfig}
\usepackage[percent]{overpic}
\usepackage{rotating}
\usepackage{pdflscape}
\usepackage{bm}
\usepackage{ragged2e}
\usepackage{booktabs}

\newcommand*{\myfont}{\fontfamily{phv}\selectfont}

\draftfalse

\journalname{JGR: Solid Earth}

\begin{document}

%
%


\title{The Episodically Buckling and Collapsing Continental Crust in Subduction Zones}

%
%




\authors{Jyoti Behura\affil{1,2}, Shayan Mehrani\affil{2}, and Farnoush Forghani\affil{3}}


\affiliation{1}{Seismic Science LLC}
\affiliation{2}{Colorado School of Mines}
\affiliation{3}{University of Colorado}




\correspondingauthor{Jyoti Behura}{jbehura@mines.edu}




\justify

\begin{keypoints}
\item A non-linear inverse correlation exists between the inter-tremor time interval and the slenderness ratio of the overriding plate in subduction zones all over the world
\item By employing all existing observations, we develop an Episodic Buckling and Collapse model of the subduction zone, wherein the overriding continental crust buckles upwards and landwards because of compressive stress from the subducting slab, and then collapses on the slab as fluid pressure in the LVZ is released
\item Geodetic measurements, previously inferred as slow slip, are the surficial expressions of slowly-evolving buckling, while the relatively short-lived tremor results from the striking of the rapidly collapsing overriding plate on the subducting slab
\item In addition to demonstrating how the model explains all existing observations and findings, we present numerical simulation study of deformation and further support the presented model with field data in the form of novel multi-component GPS analysis
\item Proposed subduction zone model has major implications for forecasting of megathrust earthquakes and for magma transport from the mantle to the shallow crust
\end{keypoints}

%
%

%
%


\begin{abstract}
We discover a remarkable correlation between the inter-tremor time interval and the slenderness ratio of the overriding plate in subduction zones all over the world. In order to understand this phenomenon better, we perform numerical simulations of 3D deformation. The numerical buckling studies show that critical load and slenderness ratio indeed have an inverse nonlinear relation between them -- identical to the classical Euler’s critical load relation, and closely resemble the relationship observed between the inter-tremor time interval and the slenderness ratio of the overriding plate. From the above analysis, we conclude that the observed relation is the result of buckling of the overriding continental plate. In addition to the above numerical analysis, we analyze the surficial 3D spatio-temporal displacements of the overriding plates in Cascadia and Alaska using 3-component GPS data. We find that these deformations are consistent with the buckling of the overriding continental crust. Based on these novel observations and guided by numerous existing scientific observations and findings, we propose an Episodic Buckling and Collapse model of subduction zones, wherein periodic geodetic changes and tectonic-tremor activity, result from the episodic buckling of the overriding continental crust and its rapid collapse on the subducting oceanic slab. According to this model, geodetic measurements, previously inferred as slow slip, are the surficial expressions of slowly-evolving buckling and rapid collapse of the overriding plate, while tremor swarms result from the striking of the collapsing overriding plate on the subducting slab (as opposed to slipping or shearing).
\end{abstract}

\section*{Plain Language Summary}
Nearly a couple of decades ago, geoscientists discovered interesting deep seismic events in subduction zones (which they termed tectonic tremor) and found that these phenomena had a strong spatio-temporal correlation with surficial displacements. This remarkable spatio-temporal correlation led them to postulate the slow-slip hypothesis wherein a part of the continental-oceanic interface shear slowly over a few days or weeks (as opposed to conventional earthquakes that span a few seconds). However, numerous observations and findings are inadequately explained by the slow-slip hypothesis. We employ all existing observations and research to develop the Episodic Buckling and Collapse model of the subduction process. We show that tremor and surficial displacements, previously associated with so-called ``slow slip'', in fact result from the episodic buckling of the overriding continental crust and its rapid collapse on the subducting oceanic slab.

%
%

%


%
%
%
%

\section{The Buckling Continental Wedge}
\begin{figure}
	\center
	\includegraphics[width=0.8\textwidth]{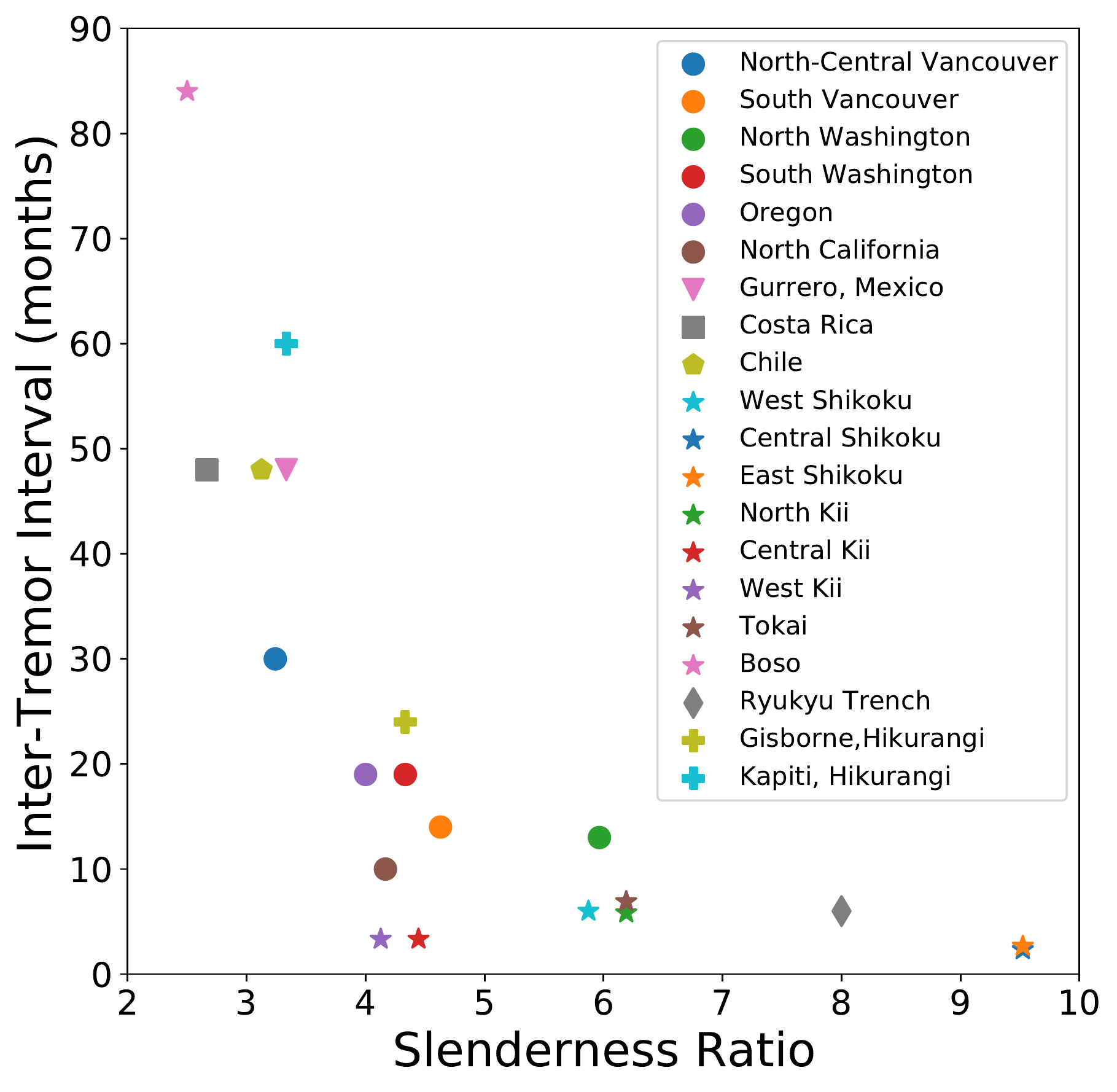}
	\caption{\textbf{Inter-tremor time interval and the slenderness ratio of continental wedges observed in subduction zones all over the world.} Data used in this plot comes from Cascadia: \protect\citeA{Kao2009,Wech2009,Beroza2011,Obara2011a}; Nankai: \protect\citeA{Beroza2011,Ozawa2007,Obara2011a}; Ryukyu: \protect\citeA{Arai2016,Obara2011a}; Gurrero, Mexico: \protect\citeA{Beroza2011,Obara2011a}; Costa Rica: \protect\citeA{Outerbridge2010,Beroza2011}; Chile: \protect\citeA{PastenAraya2018,Klein2018}; Hikurangi: \protect\citeA{Obara2011a,Wallace2010}. The slenderness ratio of a continental wedge is computed by taking the multiplicative inverse of the average plate interface slope between the depth contours of 10 and 40 km.}
	\label{fig:SlendernessRatio_Data}
\end{figure}

We observe a remarkable relationship between the inter-tremor time interval and the slenderness ratio of the overriding plate in subduction zones all over the world.
Figure~\ref{fig:SlendernessRatio_Data} shows this inverse nonlinear relationship between inter-tremor time interval and slenderness ratio for continental wedges in various subduction zones.
The inter-tremor time interval is the average time interval observed between tremor episodes at different subduction zones.
Slenderness ratio of a continental wedge is defined as the ratio of the length of the wedge to its maximum thickness, which is also equal to the inverse of the slope of the plate interface in the subduction zone.

In order to understand this interesting phenomenon, we perform numerical simulations of overriding plate deformation as well as analyze the 3D surficial deformation of the overriding continental crust in Cascadia and Alaska using GPS data.

Based on the above observations, numerical modeling of deformation, and GPS field data analysis, we present a Episodic Buckling and Collapse model of the subduction process where the continental wedge episodically buckles and collapses.
According to this model, geodetic observations previously interpreted as slow slip, are a surface manifestation of the buckling of the overriding continental crust and its subsequent rapid collapse on top of the subducting oceanic slab.
The said collapse-related striking of the continental crust on the subducting slab results in tremors and the collapse itself shows up as rapid reversals in the horizontal GPS component.
The proposed subduction model has significant and direct implication for forecasting of megathrust earthquakes and provides a `breathing' mechanism for the upwelling and flow of magmatic fluids from the upper mantle to the shallow crust.
A preliminary version of this model was initially proposed in \citeA{Behura2018} and has been modified here.

\section{Numerical Modeling of Continental Wedge Deformation}
\subsection{Euler Buckling}
Under compressive stresses slender beams spontaneously bend to form curved shapes \cite{Timoshenko1961,Gere2012}.
When the applied stress exceeds the yield strength, the material experiences an irreversible plastic or brittle deformation.
\emph{Buckling, on the other hand, occurs at stresses much lower than the yield strength of the structure \cite{Timoshenko1961,Gere2012}.}
The buckling of beams is determined by the material's Young's modulus and its slenderness.
Also, the more slender the structure, lower is the critical stress needed for buckling.
Slenderness is a measure of the tendency of the beam to buckle and is quantified by the slenderness ratio -- the ratio between the effective length of the beam and its radius of gyration \cite{Timoshenko1961,Gere2012,Eisley2011}.
For a given slenderness ratio, there is a critical load (lower than the yield stress of the material), at which the wedge will bend (buckling/folding) before it can break.

Geological fold structures, that form under various stress and pressure conditions and are observed at a wide range of scales, are a prime example of buckling.

\begin{figure}
	\center
	\includegraphics[width=0.8\textwidth]{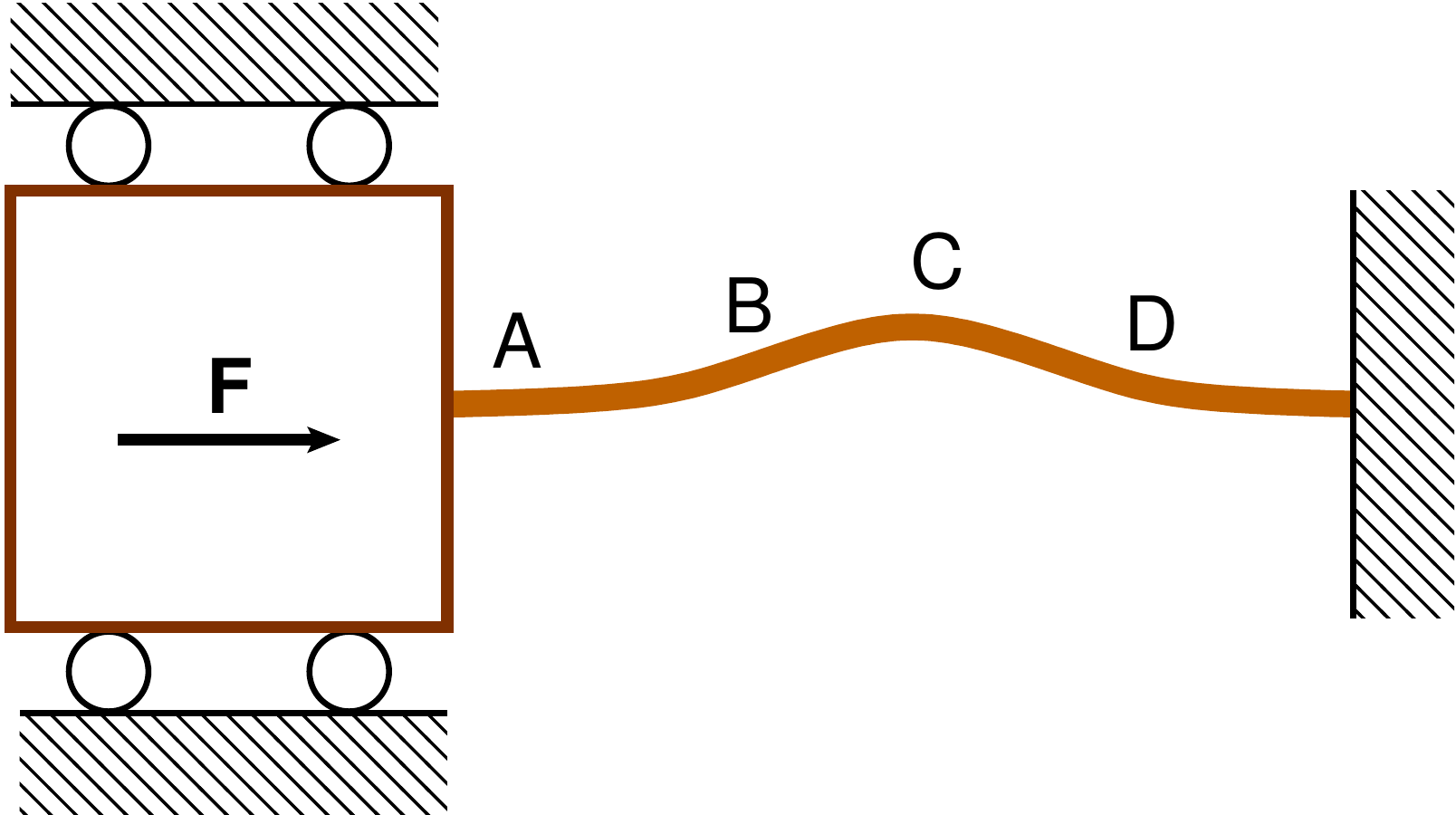}
	\caption{\textbf{Schematic of Euler buckling mode with both ends fixed.} Stress {\myfont \textbf{F}} is applied by the subducting slab at the locked zone (seaward fixed end). The landward fixed end results from the immovable back-arc continental crust. Locations A, B, C, and D, correspond to the positions on the continental crust with their net displacement analyzed later in Figure~\ref{fig:GPSandHodo}.}
	\label{fig:BucklingCartoon}
\end{figure}

In subduction zones, the overriding continental crust may be considered as a collection of trench-perpendicular slender beams (because of the plane stress imposed by the subducting slab) and may buckle under the compressive stress applied by the subducting slab at the locked zone.
Given that the average compressive stress exerted by spreading ridges is approximately 25~MPa \cite{Solomon1980} and the average yield strength of continental lithosphere is close to 400~MPa \cite{Brace1980,Burov2011}, we expect buckling to be the predominant deformation mechanism (instead of plastic or brittle behavior).
A schematic scenario of forces and boundary conditions experienced by the continental crust is shown in Figure~\ref{fig:BucklingCartoon}.
The seaward locked zone and the landward thick continental crust would result in deformation akin to the Euler buckling mode where both ends are fixed \cite{Timoshenko1961,Gere2012}.
The seaward end, however, can slide because of the landward movement of the oceanic crust (Figure~\ref{fig:BucklingCartoon}).
Such a buckling mode results in not only horizontal displacements but also significant vertical strain in the continental crust.

In order to ascertain and verify the above intuition, we perform numerical simulation of the static deformation of wedges by subjecting them to similar forces and boundary conditions encountered at subduction zones.

\subsection{Numerical Simulation of Buckling} \label{Buckling}
\begin{figure}
	\center
	\subfloat[Model]{\label{fig:WedgeModel}\includegraphics[width=\textwidth]{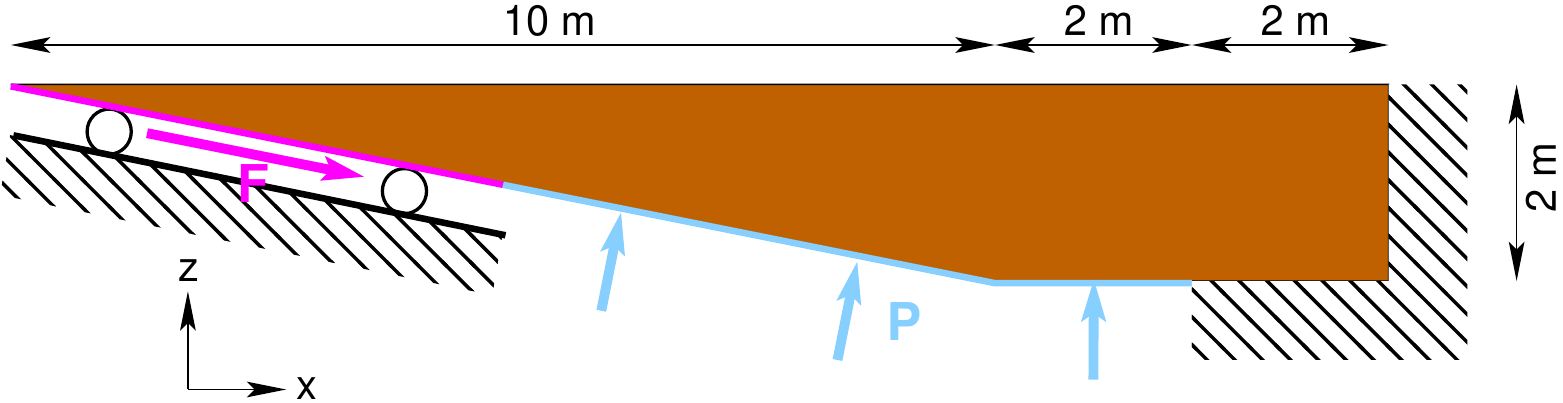}}\hfill\null\\
	\subfloat[X (horizontal) displacement]{\label{fig:WedgeModel_UX}\includegraphics[width=0.9\textwidth,trim=33mm 93mm 30mm 85mm,clip=true]{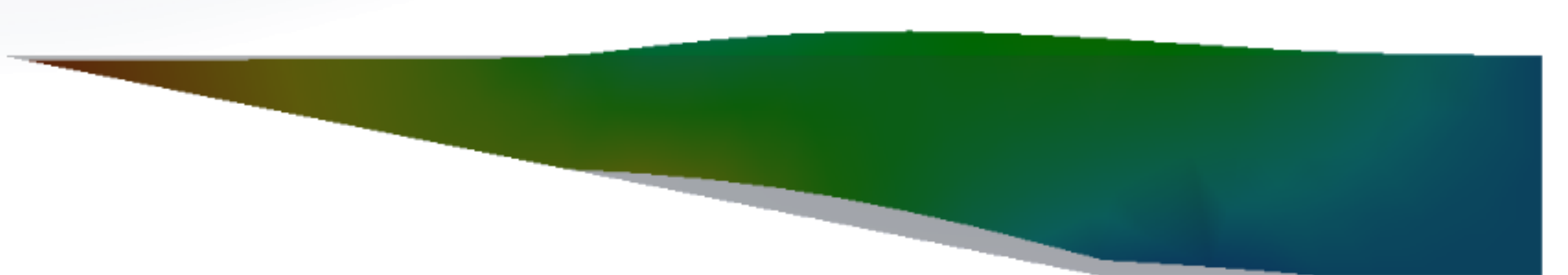}}\hfill
	\subfloat{\includegraphics[width=0.095\textwidth]{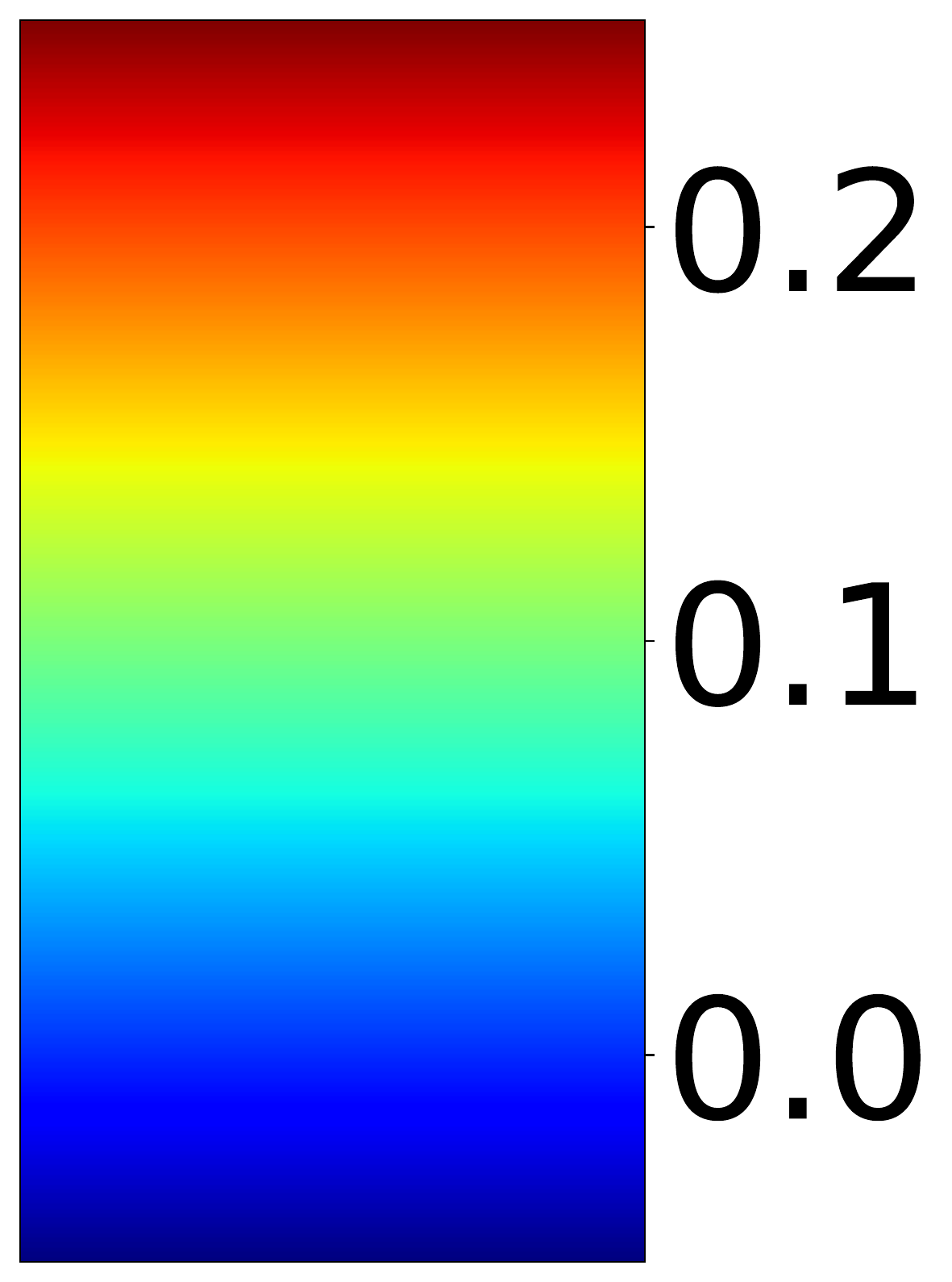}}\\
	\addtocounter{subfigure}{-1}
	\subfloat[Z (vertical) displacement]{\label{fig:WedgeModel_UZ}\includegraphics[width=0.9\textwidth,trim=33mm 93mm 30mm 85mm,clip=true]{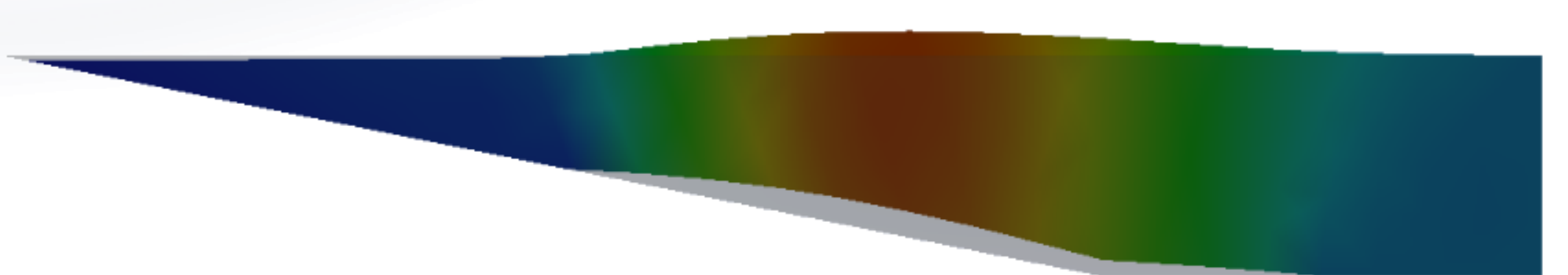}}\hfill
	\subfloat{\includegraphics[width=0.095\textwidth]{Colormap.pdf}}\\
	\addtocounter{subfigure}{-1}
	\subfloat[Resultant displacement]{\label{fig:WedgeModel_U}\includegraphics[width=0.9\textwidth,trim=33mm 93mm 30mm 85mm,clip=true]{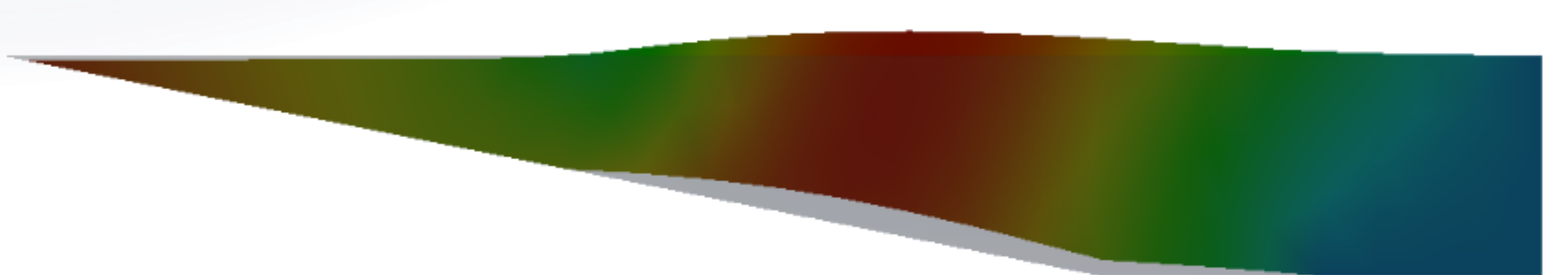}}\hfill
	\subfloat{\includegraphics[width=0.095\textwidth]{Colormap.pdf}}
	\caption{\textbf{(a) Wedge-model (akin to the overriding continental wedge) used in simulation of static deformation.} {\myfont \textbf{F}} corresponds to the shear stress applied by the subducting slab at the locked zone (magenta line). {\myfont \textbf{P}} represents the normal stress exerted by the near-lithostatic pore-fluid pressure. Only sliding (displacement along the plate interface) is allowed in the locked zone, while the landward edge remains fixed. All other surfaces have free boundary conditions. \textbf{Horizontal (b), vertical (c), and net (d) displacements (in meters) superimposed on the deformed wedge resulting from the forces and boundary conditions in (a).} The undeformed shape, in the form of a transparent body, is also superimposed. The colormap for the displacements ranges from -0.05 m to 0.25 m.}
	\label{fig:StaticModeling}
\end{figure}

Here, we simulate the deformation of a wedge-shaped body (akin to the overriding continental wedge) subjected to stresses and boundary conditions encountered by the continental crust in subduction zones as shown in Figure~\ref{fig:StaticModeling}.
All numerical simulations are carried out in Solidworks \cite{Solidworks}.

Computational expense and software constraints limit us to performing numerical simulations on small-scale 2-dimensional models (Figure~\ref{fig:WedgeModel}) instead of models on the order of tens of kilometers and in 3-dimensions.
Ideally, because subduction zone structures are 3-dimensional, one should be simulating continental crust deformation using plate buckling (3D) \cite{Timoshenko1961,Gere2012}, instead of column buckling (2D) to accurately explain 3D deformation and local buckling.
Also, to perform a more complicated quantitative analysis, one will have to perform simulations at the true scale, use a 3D heteregeneous viscoelastic earth model, and impose accurate stresses and precise boundary conditions.
Nonetheless, the findings are equally applicable to the same geometry at large scales, especially for qualitative analysis of deformation.

For the wedge material, we use plain carbon steel having a modulus of elasticity of 200~GPa.
The landward edge of the wedge is assumed to be immovable (Figure~\ref{fig:WedgeModel}) (zero displacement).
In the locked zone, only sliding along the interface is allowed with the interface-normal displacement set to zero.
All other surfaces are free boundaries.
Both shear stress {\myfont \textbf{F}} and pore-pressure {\myfont \textbf{P}} are set at 1~GPa for the static deformation (in the subsequent section, we also perform simulations in the absence of any pore-pressure to analyze its role).

It is clear from Figure~\ref{fig:StaticModeling}, that the wedge buckles under the forces and boundary conditions imposed on it.

The horizontal displacement (Figure~\ref{fig:WedgeModel_UX}) is maximum at the seaward edge and decreases monotonically away from it.
Vertical displacement, shown in Figure~\ref{fig:WedgeModel_UZ}, is significantly large in the middle between the locked zone and the fixed landward edge.
Below, in section~\ref{GPSAnalysis}, we show how the horizontal and vertical GPS measurements in Cascadia and Alaska closely correspond to the surficial displacement patterns seen in Figures~\ref{fig:WedgeModel_UX} and \ref{fig:WedgeModel_UZ}.

To decipher the effect of the wedge geometry on buckling, we perform numerical simulation of the buckling process of on a range of wedge geometries similar to the one shown in Figure~\ref{fig:WedgeModel}.
The forces and boundary conditions are the same as shown in Figure~\ref{fig:WedgeModel}, with the only difference being in the geometry.
For a wedge geometry, we redefine the slenderness ratio to be the ratio of the length of the wedge to the maximum thickness, which is also equal to the inverse of the slope of the plate interface in subduction zones.
For example, the slenderness ratio of the wedge in Figure~\ref{fig:WedgeModel} is $10/2=5$.

\begin{figure}
	\center
	\includegraphics[width=0.8\textwidth]{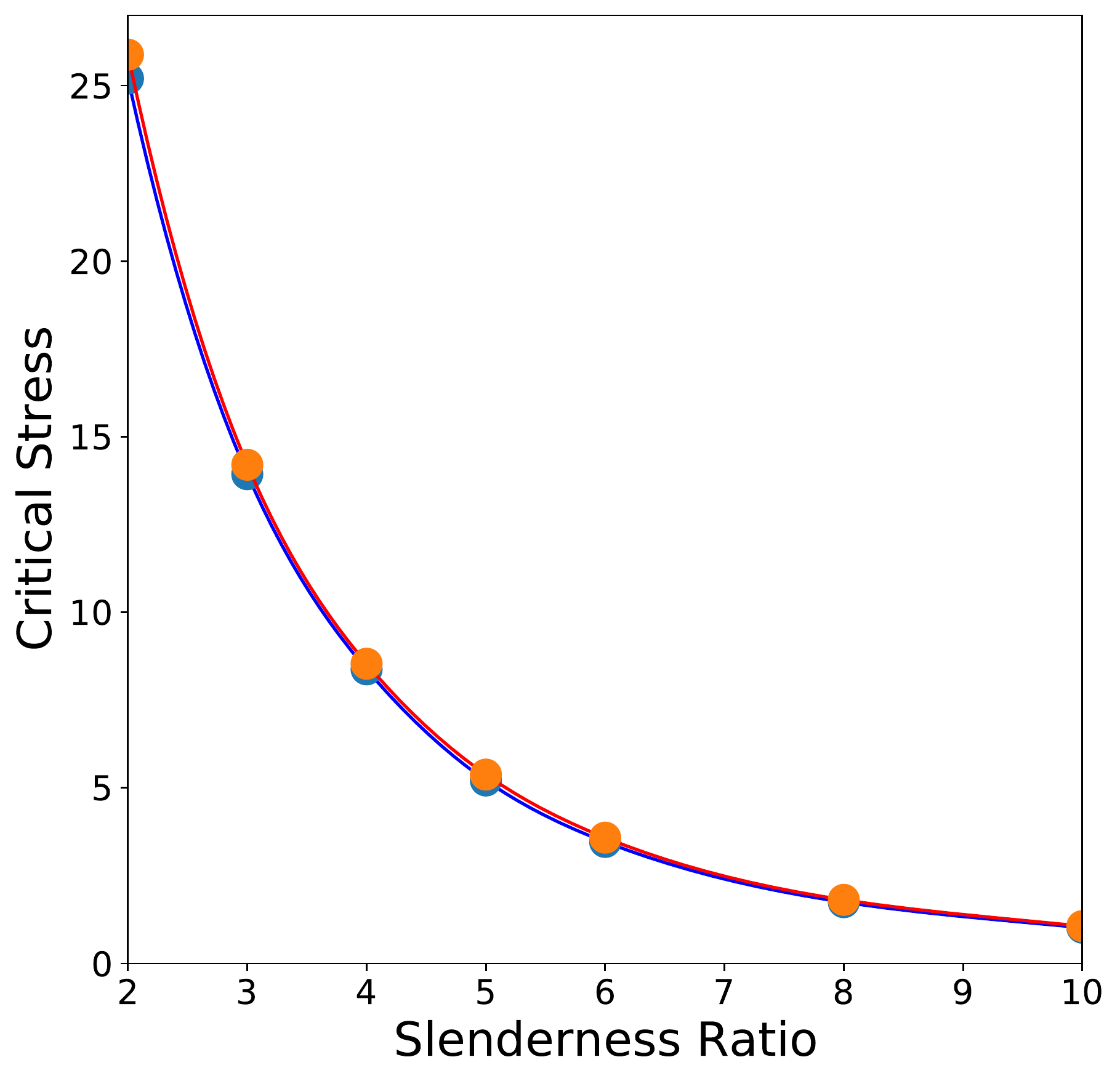}
	\caption{\textbf{Critical stress (in GPa) as a function of slenderness ratio for a wedge experiencing force {\myfont\textbf{F}} and boundary conditions shown in Figure~\ref{fig:WedgeModel} in the presence (blue line) and absence (orange line) of pore-fluid pressure {\myfont\textbf{P}}.} The dots represent values computed numerically for a range on geometries and the lines are obtained using interpolation through these points. Note the similarity with the trend in Figure~\ref{fig:SlendernessRatio_Data}.}
	\label{fig:SlendernessRatio_Modeling}
\end{figure}

We perform numerical buckling simulation of wedges for a range of slenderness ratios in Solidworks \cite{Solidworks} with and without pore-pressure {\myfont\textbf{P}} and compute the critical stress for each scenario.
The results (Figure~\ref{fig:SlendernessRatio_Modeling}) show that critical load and slenderness ratio have an inverse nonlinear relation between them and this relation is quite similar to the classical Euler's critical load relation for beams \cite{Timoshenko1961,Gere2012,Eisley2011}.
The above numerical analysis confirms that wedges subject to forces and boundary conditions akin to those encountered at subduction zones will experience buckling.

\subsection{Field Data}
Although the trends of curves in Figures~\ref{fig:SlendernessRatio_Data} and \ref{fig:SlendernessRatio_Modeling} are similar, they are not amenable to a direct comparison unless we can compute the slenderness ratios of continental wedges, and more importantly establish a relation between inter-tremor time interval and critical stress.

Slenderness ratio of a overriding continental crust is equal to the multiplicative inverse of the slope of the plate interface, and is therefore quite straightforward to compute from imaging studies carried out at different subduction zones.

Estimation of critical stress, however, is much more challenging.
Therefore, we resort to an indirect measure of the critical load for continental wedges around the world.
Tremors (and so-called slow slip events) display a wide range of periodicity in the Nankai and Hikurangi subduction zones \cite{Schwartz2007,Wallace2010,Obara2011a} -- with the seismicity characteristics clearly correlated to the depth of the seismicity \cite{Wallace2010}.
A thinner crust is more easily buckled than a thicker one -- it will take lesser time and lesser force to buckle; at the same time, the thinner crust can accommodate a lesser degree of strain energy than a thicker one.
In sections \ref{GPSAnalysis} and \ref{EBC} below, we show that buckling of the continental plate (and its subsequent collapse) happen periodically.
Hence, a thinner crust will undergo more cycles of buckling (and collapse) than a thicker one within the same time period, all the while releasing lesser seismic energy in each cycle.
In other words, the inter-tremor time interval for thinner crusts (high slenderness ratio) will be smaller than for thick crusts (low slenderness ratio).
If we assume that all continental wedges experience the same stress rate, then the critical stress should be approximately directly proportional to the inter-tremor time interval (because stress equals the product of stress rate and time).
\emph{Therefore, we use inter-tremor time interval as a proxy for critical stress.}

Under the light of the above argument and given the close correspondence between the curves in Figures~\ref{fig:SlendernessRatio_Data} and \ref{fig:SlendernessRatio_Modeling}, \emph{we conclude that forces and boundary conditions prevalent at subduction zones result in buckling of the overriding continental plate.}

One might argue that static friction might explain the observation in Figure~\ref{fig:SlendernessRatio_Data}.
However, assuming a constant coefficient of friction across all plate interfaces and a similar continental crust thickness in all subduction zones, static friction will result in a positive correlation between inter-tremor time interval and the slenderness ratio.
That is because, with increasing slenderness ratio (which corresponds to decreasing dip of the plate interface), the normal force on the plate interface increases, thereby thereby increasing the frictional force -- which would translate to an approximately linearly increasing inter-tremor interval time with increasing slenderness ratio.
However, we observe the opposite relationship in subduction zones as shown in Figure~\ref{fig:SlendernessRatio_Data}), which rules out friction as an explanation for this phenomenon.

Also, it is also clear from Figure~\ref{fig:SlendernessRatio_Modeling} that the effect of the pore-fluid pressure on the buckling propensity is low and the primary force controlling the buckling process is the shear stress applied by the subducting slab in the locked zone.

Note that subduction zone geometries and rheological properties are complicated, vary spatially, and evolve with time, which explains the scatter and presence of outliers in Figure~\ref{fig:SlendernessRatio_Data}.
The slenderness ratio used in Figure~\ref{fig:SlendernessRatio_Data} for any region is computed by averaging the values along dip as well as strike.
Imaging studies show that there may be significant variability in the slope of a plate interface.
Also, in addition to the subduction rate, the inter-tremor time interval is influenced by other factors, the most important of which include 3D geometry of the overriding plate, modulus of elasticity of the overriding plate, and heterogeneities therein.
Prime examples of such complicated subduction zones include Alaska and Japan which exhibit a range of inter-tremor time intervals.

\section{Spatio-Temporal Surface Deformation in Cascadia and Alaska}\label{GPSAnalysis}
If buckling of the overriding continental crust is indeed occurring, one should be able to observe it on surface displacement data.
To ascertain this, we analyze all three components of GPS recording (two horizontal and one vertical) in Cascadia and Alaska to study the 3D deformation of the continental crust as a function of space.

Equally importantly, buckling of continental crusts should have a temporal signature.
In other words, a structure undergoing buckling should demonstrate vertical and horizontal displacements that are a function of time.
Buckling, however, cannot go on forever.
If the strain rate remains constant, the structure will experience pastic deformation.
On the other hand, if the stresses are reduced or eliminated, the structure will revert to its original state as the initial buckling is elastic.

Although we observe the overriding plate in subduction zones is buckling, we do not see any large-scale permanent folding structures that would point to plastic deformation.
Moreover, as mentioned above, the average compressive stress exerted by spreading ridges is approximately 25~MPa \cite{Solomon1980} and the average yield strength of continental lithosphere is close to 400~MPa \cite{Brace1980,Burov2011}, which implies that one should expect buckling to be the predominant deformation mechanism (instead of plastic or brittle behavior) of the overriding continental wedge.
\emph{Therefore, it is resonable to conclude that the stresses causing buckling of continental plates are getting reduced or eliminated at certain times -- resulting in a collapse of the continental plate.}

Below, we show that geodetic observations previously interpreted as slow slip, are in fact a surface manifestation of the buckling of the overriding continental crust and its subsequent rapid collapse on top of the subducting oceanic slab.
We also demonstrate that the buckling and collapse of the overriding plate occurs periodically at subduction zones.

\subsection{Displacements due to Buckling and Collapse}
\begin{figure}
	\center
	\subfloat{
		\label{fig:A_GPS}
		\begin{overpic}[width=0.7\textwidth]{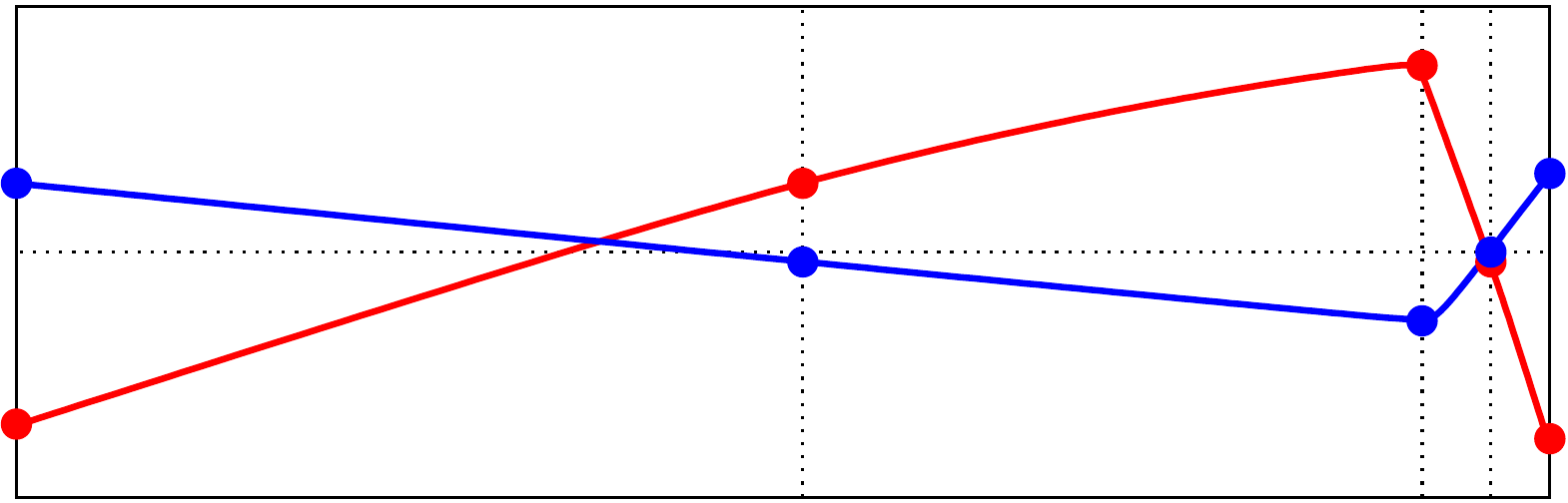}
			\put(-3,15){\rotatebox[origin=c]{90}{{\myfont 0}}}
			\put(-3,7){\rotatebox[origin=c]{90}{{\myfont -ve}}}
			\put(-3,22){\rotatebox[origin=c]{90}{{\myfont +ve}}}
			\put(-7,15){\rotatebox[origin=c]{90}{{\myfont Displacement}}}
			\linethickness{0.7mm}
			\put(4,28){\myfont (A)}
			\put(70,6){\color{red}\line(1,0){5}}
			\put(76,5){\color{red}{\myfont X}}
			\put(70,3){\color{blue}\line(1,0){5}}
			\put(76,2){\color{blue}{\myfont Z}}
		\end{overpic}
	}
	\subfloat{
		\label{fig:A_Hodo}
		\begin{overpic}[width=0.225\textwidth]{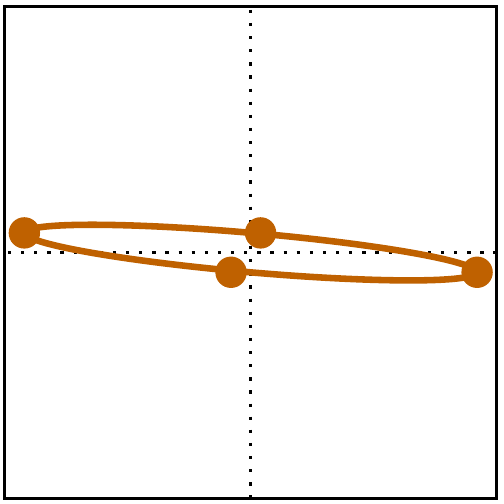}
			\put(105,47){\color{blue}\scalebox{1}{\rotatebox[origin=c]{90}{{\myfont 0}}}}
			\put(115,47){\color{blue}\scalebox{1}{\rotatebox[origin=c]{90}{{\myfont Z}}}}
			\put(105,20){\color{blue}\scalebox{1}{\rotatebox[origin=c]{90}{{\myfont -ve}}}}
			\put(105,69){\color{blue}\scalebox{1}{\rotatebox[origin=c]{90}{{\myfont +ve}}}}
			\put(5,60){\myfont \textbf{T}$\bm{_0}$,\textbf{T}$\bm{_6}$}
			\put(55,55){\myfont \textbf{T}$\bm{_1}$}
			\put(87,30){\myfont \textbf{T}$\bm{_2}$}
			\put(40,35){\myfont \textbf{T}$\bm{_4}$}
			\put(4,88){\myfont (A)}
		\end{overpic}
	}\\
	\subfloat{
		\label{fig:B_GPS}
		\begin{overpic}[width=0.7\textwidth]{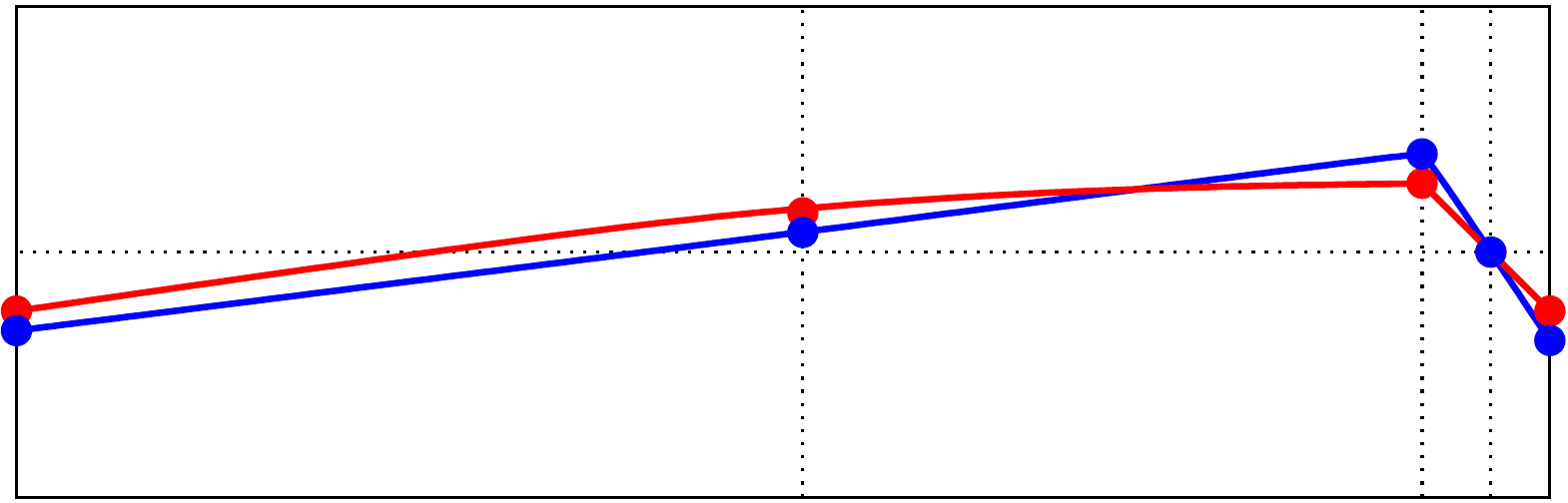}
			\put(-3,15){\rotatebox[origin=c]{90}{{\myfont 0}}}
			\put(-3,7){\rotatebox[origin=c]{90}{{\myfont -ve}}}
			\put(-3,22){\rotatebox[origin=c]{90}{{\myfont +ve}}}
			\put(-7,15){\rotatebox[origin=c]{90}{{\myfont Displacement}}}
			\linethickness{0.7mm}
			\put(4,28){\myfont (B)}
			\put(70,6){\color{red}\line(1,0){5}}
			\put(76,5){\color{red}{\myfont X}}
			\put(70,3){\color{blue}\line(1,0){5}}
			\put(76,2){\color{blue}{\myfont Z}}
		\end{overpic}
	}
	\subfloat{
		\label{fig:B_Hodo}
		\begin{overpic}[width=0.225\textwidth]{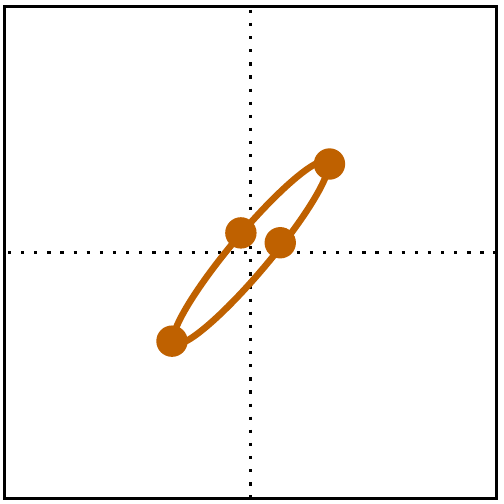}
			\put(105,47){\color{blue}\scalebox{1}{\rotatebox[origin=c]{90}{{\myfont 0}}}}
			\put(115,47){\color{blue}\scalebox{1}{\rotatebox[origin=c]{90}{{\myfont Z}}}}
			\put(105,20){\color{blue}\scalebox{1}{\rotatebox[origin=c]{90}{{\myfont -ve}}}}
			\put(105,69){\color{blue}\scalebox{1}{\rotatebox[origin=c]{90}{{\myfont +ve}}}}
			\put(10,35){\myfont \textbf{T}$\bm{_0}$,\textbf{T}$\bm{_6}$}
			\put(60,41){\myfont \textbf{T}$\bm{_1}$}
			\put(70,64){\myfont \textbf{T}$\bm{_2}$}
			\put(35,58){\myfont \textbf{T}$\bm{_4}$}
			\put(4,88){\myfont (B)}
		\end{overpic}
	}\\
	\subfloat{
		\label{fig:C_GPS}
		\begin{overpic}[width=0.7\textwidth]{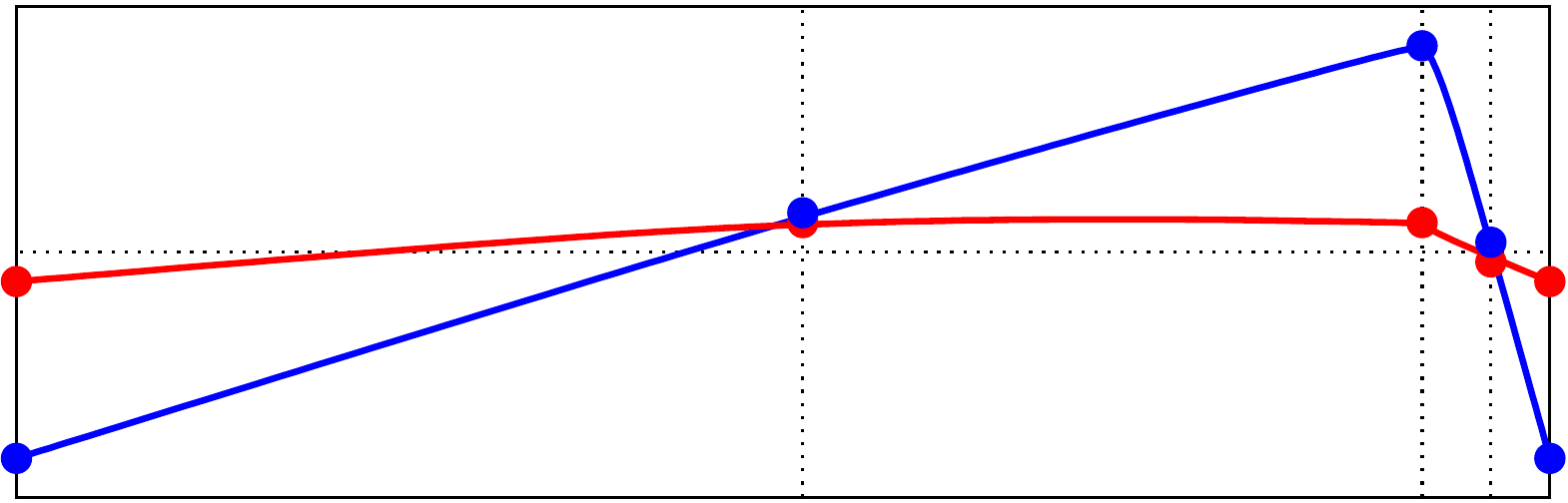}
			\put(-3,15){\rotatebox[origin=c]{90}{{\myfont 0}}}
			\put(-3,7){\rotatebox[origin=c]{90}{{\myfont -ve}}}
			\put(-3,22){\rotatebox[origin=c]{90}{{\myfont +ve}}}
			\put(-7,15){\rotatebox[origin=c]{90}{{\myfont Displacement}}}
			\linethickness{0.7mm}
			\put(4,28){\myfont (C)}
			\put(70,6){\color{red}\line(1,0){5}}
			\put(76,5){\color{red}{\myfont X}}
			\put(70,3){\color{blue}\line(1,0){5}}
			\put(76,2){\color{blue}{\myfont Z}}
		\end{overpic}
	}
	\subfloat{
		\label{fig:C_Hodo}
		\begin{overpic}[width=0.225\textwidth]{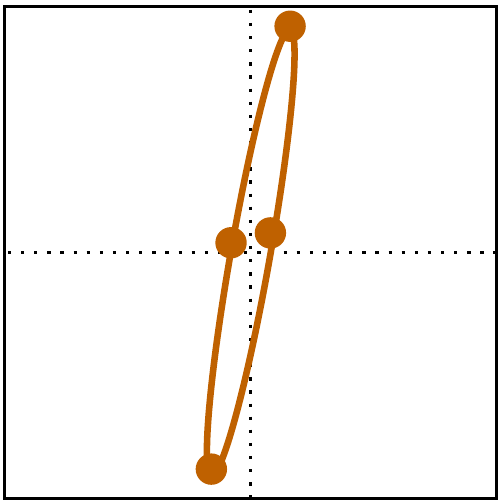}
			\put(105,47){\color{blue}\scalebox{1}{\rotatebox[origin=c]{90}{{\myfont 0}}}}
			\put(115,47){\color{blue}\scalebox{1}{\rotatebox[origin=c]{90}{{\myfont Z}}}}
			\put(105,20){\color{blue}\scalebox{1}{\rotatebox[origin=c]{90}{{\myfont -ve}}}}
			\put(105,69){\color{blue}\scalebox{1}{\rotatebox[origin=c]{90}{{\myfont +ve}}}}
			\put(15,5){\myfont \textbf{T}$\bm{_0}$,\textbf{T}$\bm{_6}$}
			\put(60,52){\myfont \textbf{T}$\bm{_1}$}
			\put(62,90){\myfont \textbf{T}$\bm{_2}$}
			\put(30,55){\myfont \textbf{T}$\bm{_4}$}
			\put(4,88){\myfont (C)}
		\end{overpic}
	}\\
	\subfloat{
		\label{fig:D_GPS}
		\begin{overpic}[width=0.7\textwidth]{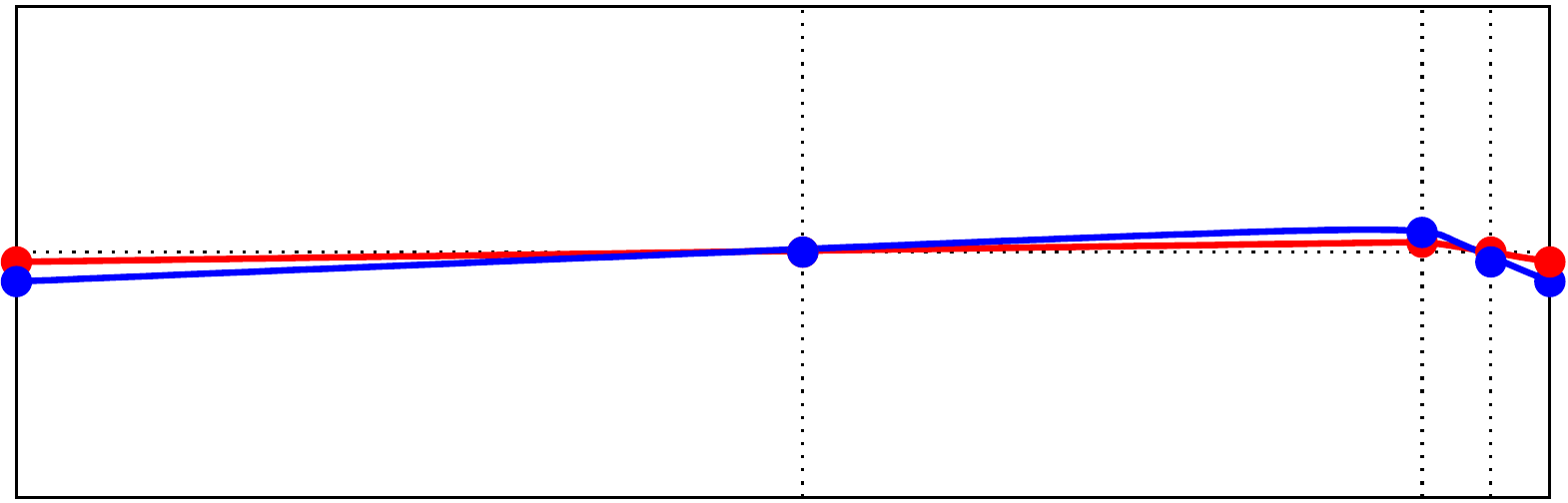}
			\put(-3,15){\rotatebox[origin=c]{90}{{\myfont 0}}}
			\put(-3,7){\rotatebox[origin=c]{90}{{\myfont -ve}}}
			\put(-3,22){\rotatebox[origin=c]{90}{{\myfont +ve}}}
			\put(-7,15){\rotatebox[origin=c]{90}{{\myfont Displacement}}}
			\linethickness{0.7mm}
			\put(4,28){\myfont (D)}
			\put(70,6){\color{red}\line(1,0){5}}
			\put(76,5){\color{red}{\myfont X}}
			\put(70,3){\color{blue}\line(1,0){5}}
			\put(76,2){\color{blue}{\myfont Z}}
			\put(0,-3){\myfont \textbf{T}$\bm{_0}$}
			\put(50,-3){\myfont \textbf{T}$\bm{_1}$}
			\put(90,-3){\myfont \textbf{T}$\bm{_2}$}
			\put(94,-3){\myfont \textbf{T}$\bm{_4}$}
			\put(98,-3){\myfont \textbf{T}$\bm{_6}$}
		\end{overpic}
	}
	\subfloat{
		\label{fig:D_Hodo}
		\begin{overpic}[width=0.225\textwidth]{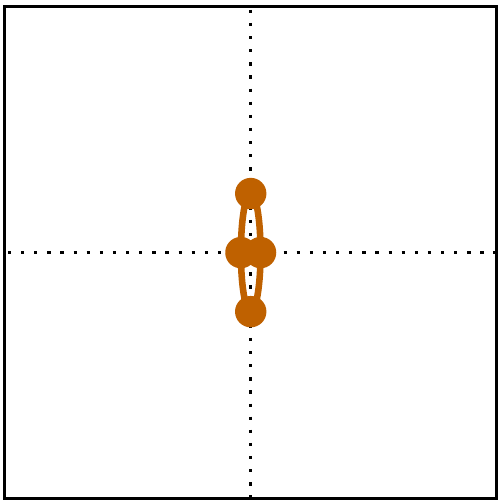}
			\put(48,-10){\color{red}\scalebox{1}{{\myfont 0}}}
			\put(47,-20){\color{red}\scalebox{1}{{\myfont X}}}
			\put(20,-10){\color{red}\scalebox{1}{{\myfont -ve}}}
			\put(69,-10){\color{red}\scalebox{1}{{\myfont +ve}}}
			\put(105,47){\color{blue}\scalebox{1}{\rotatebox[origin=c]{90}{{\myfont 0}}}}
			\put(115,47){\color{blue}\scalebox{1}{\rotatebox[origin=c]{90}{{\myfont Z}}}}
			\put(105,20){\color{blue}\scalebox{1}{\rotatebox[origin=c]{90}{{\myfont -ve}}}}
			\put(105,69){\color{blue}\scalebox{1}{\rotatebox[origin=c]{90}{{\myfont +ve}}}}
			\put(40,25){\myfont \textbf{T}$\bm{_0}$,\textbf{T}$\bm{_6}$}
			\put(58,48){\myfont \textbf{T}$\bm{_1}$}
			\put(45,68){\myfont \textbf{T}$\bm{_2}$}
			\put(35,48){\myfont \textbf{T}$\bm{_4}$}
			\put(4,88){\myfont (D)}
		\end{overpic}
	}
	\vspace{8mm}
	\caption{\textbf{Schematic time-dependent detrended displacements (left column) and corresponding hodograms (right column) of points A through D (Figure~\ref{fig:BucklingCartoon}) during a single cycle of Episodic Buckling and Collapse.} Horizontal displacement X is shown in red and vertical displacement Z in blue. The different phases of the subduction cycle are also denoted.}
	\label{fig:GPSandHodo}
\end{figure}

As argued above, buckling of the overriding plate in subduction zones is followed by a collapse phase.
Figure~\ref{fig:GPSandHodo} shows a schematic of the expected temporal evolution of vertical (blue) and horizontal (red) displacements of four locations A, B, C, and D, (Figure~\ref{fig:BucklingCartoon}) on the surface of a continental plate through a single buckling and collapse cycle.
Spatial displacement patterns in Figure~\ref{fig:GPSandHodo} are made to be consistent with the numerical static modeling results shown in Figure~\ref{fig:StaticModeling}.
The magnitude of the horizontal displacement is expected to decrease monotonically from the corner of the accretionary wedge (location A) landward as depicted by the decreasing range of the horizontal displacement moving from A through D.
The vertical displacement, however, is small at location A, attains a maximum at location C, and tapers off to a small value further landwards (location D).

An efficient technique to analyze and quantify such multi-component data is to generate hodograms which are a display of the motion of a point as a function of time.
Figure~\ref{fig:GPSandHodo} shows the hodograms for each of the four locations A, B, C, and D on the right.
The path followed by a particle during the buckling phase is different from that followed during the collapse phase, thereby resulting in hysteresis of the particle motion.
Note that such hysteresis demonstrates a non-linear particle motion (Figure~\ref{fig:GPSandHodo}) as opposed to an linear motion (with near-zero vertical displacement) expected for the case of slow slip.
Moreover, it is clear from the hodograms that the horizontal displacement decreases monotonically from the corner of the accretionary wedge (location A) landward, while the vertical displacement attains a maximum somewhere between the locked zone and the backarc.

The tilt of the major-axis of the hodogram with respect to the vertical is also characteristic of buckling-induced displacements.
The hodogram major-axis in the vicinity of the seaward-edge of the overriding plate (location A) is close to horizontal (tilt of 90$^\circ$).
The tilt at location C, on the other hand, is close to 0$^\circ$.
In between locations A and C, the tilt is expected to systematically change from 90$^\circ$ at A to 0$^\circ$ at C.
Because of the absence of geodetic measurements in the seaward wedge of the overriding plate, however, we expect to see tilts corresponding only to locations B, C, and D.

\begin{figure}
	\center
	\subfloat[GPS]{\label{fig:ALBH_v}\includegraphics[width=\textwidth]{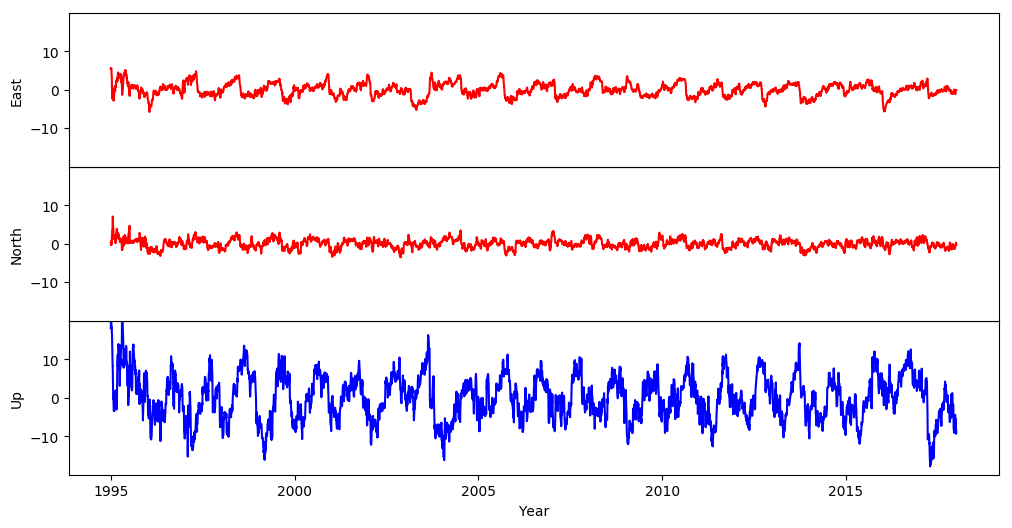}}\\
	\subfloat[Hodogram]{\label{fig:ALBH_h}\includegraphics[width=0.7\textwidth]{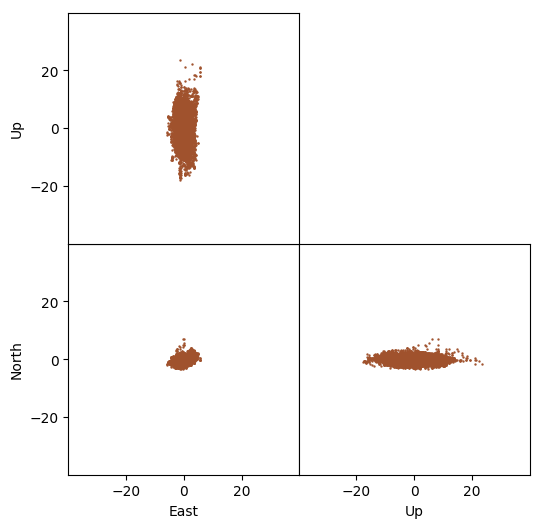}}
	\caption{\textbf{East, North, and vertical components of GPS data and corresponding hodogram from the Albert Head GPS site on Vancouver Island in Victoria, British Columbia and corresponding hodogram on the right.} All data have been detrended and filtered. The hodogram is displayed in the form of projections on the three orthogonal planes.}
	\label{fig:ALBH}
\end{figure}

Figure~\ref{fig:ALBH} shows an example of a hodogram obtained from GPS data.
This data comes from the Albert Head GPS site on Vancouver Island in Victoria, British Columbia -- the data for which was originally employed by \citeA{Rogers2003} to hypothesize the process of slow slip.
Note the hysteresis and the prominent vertical displacement observed at this site which is quite similar to the pattern expected for surface location C (Figure~\ref{fig:GPSandHodo}).
Below, in Section~\ref{EBC}, we hypothesize that location C lies right above the tremor zone.

\subsection{Vertical GPS Measurements}
Uncertainty in vertical GPS measurements is approximately 3 times that of horizontal measurements.
More importantly, we recognize that seasonal variations in surface mass variations can have substantial impact on vertical GPS measurements \cite{Blewitt2001,Dong2002,Bettinelli2008}.

Here, however, we ignore the effect of seasonal changes on vertical GPS measurements because it is extremely challenging to decouple the effect of seasonal surficial mass changes from displacement due to tectonic deformation.
This task become especially challenging in Cascadia where the episodic deformation cycle, spanning 13--14 months, is close to the seasonal cycle (12 months).

In some cases seasonal effects can be reliably accounted for using GRACE-based models \cite{Fu2013}.
GRACE-based models, however, are still not error-proof because if there is tectonic-related uplifting/collapse, there will be related gravity perturbations that will be contained in GRACE measurements (and superimposed on seasonal changes).

Still, other studies \cite{Douglas2005,Miyazaki2006,Heki2008,Wallace2010,Liu2015} have clearly shown the close correspondence between the patterns seen on vertical displacements with horizontal measurements.
More recently, \citeA{Klein2018} clearly show slow slip events on horizontal and vertical GPS measurements in Chile and further show that only tectonic processes (and not instrumental, hydrologic, oceanic, or atmospheric loading processes) could be generating such transient signals.

\subsection{Data Processing}
Prior to hodogram analysis, we detrend GPS data using a 1001 point median filter to eliminate long-term trends, and thereafter de-noise it using a 11-point median filter to suppress short-term noise bursts.
GPS stations with significant noise that could not be corrected from using the above filtering operations are not used in the analysis.

Computation of the net vertical and horizontal GPS displacements is done by fitting ellipsoids to the hodograms.
Projection of the major axis of the ellipse on the vertical direction and the horizontal plane yields the net vertical and horizontal displacements, respectively.

\subsection{3D Displacements in Alaska and Cascadia}
\begin{landscape}
\begin{figure}
	\center
	\subfloat[Vertical displacement]{\label{fig:Alaska_v}\includegraphics[height=0.48\textheight]{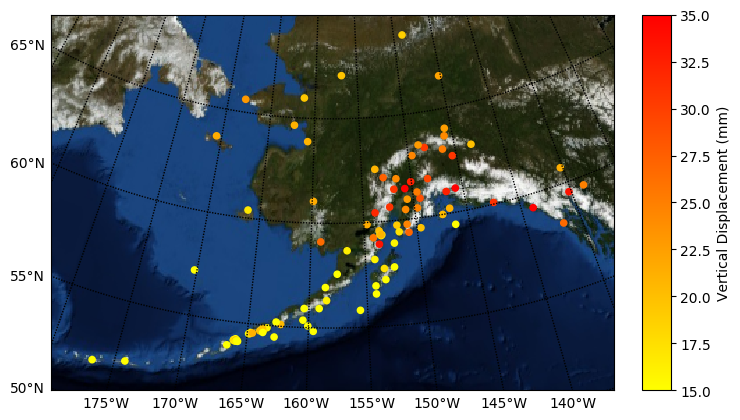}}\hfill
	\subfloat[Horizontal displacement]{\label{fig:Alaska_h}\includegraphics[height=0.48\textheight]{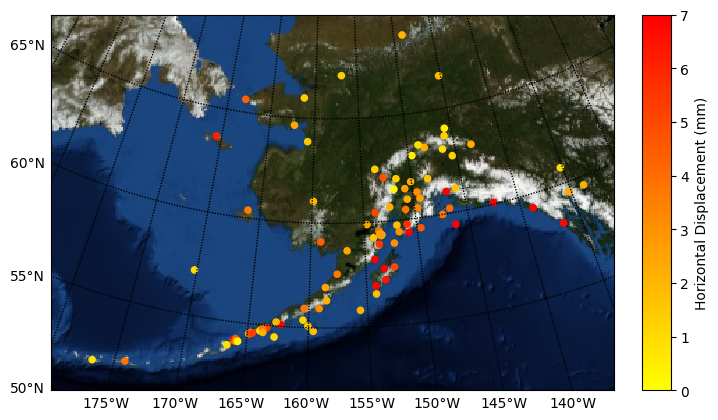}}\\
	\subfloat[Vertical-Horizontal Ratio]{\label{fig:Alaska_ratio}\includegraphics[height=0.48\textheight]{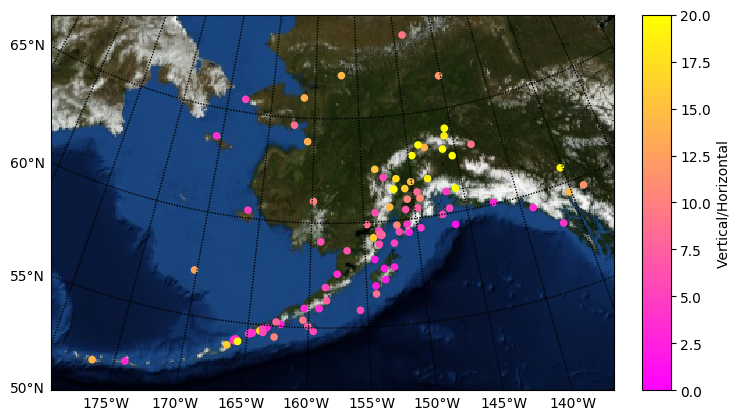}}\hfill
	\subfloat[Tilt]{\label{fig:Alaska_tilt}\includegraphics[height=0.48\textheight]{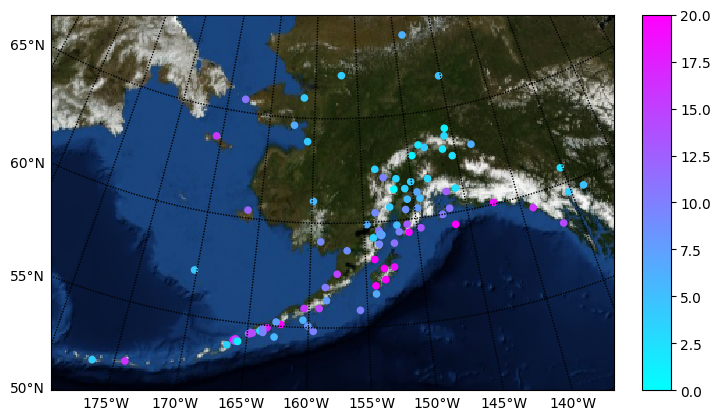}}
	\caption{\textbf{Measures of surface deformation in Alaska.} \textbf{a,} Net vertical displacement and \textbf{b,} net horizontal displacement computed from GPS measurements, \textbf{c,} their ratio, and \textbf{d,} hodogram tilt (in degrees) from the vertical. All color scales have been truncated to expose the trends.}
	\label{fig:Alaska}
\end{figure}
\end{landscape}

\begin{figure}
	\center
	\subfloat[Vertical displacement]{\label{fig:Cascadia_v}\includegraphics[height=0.35\textheight]{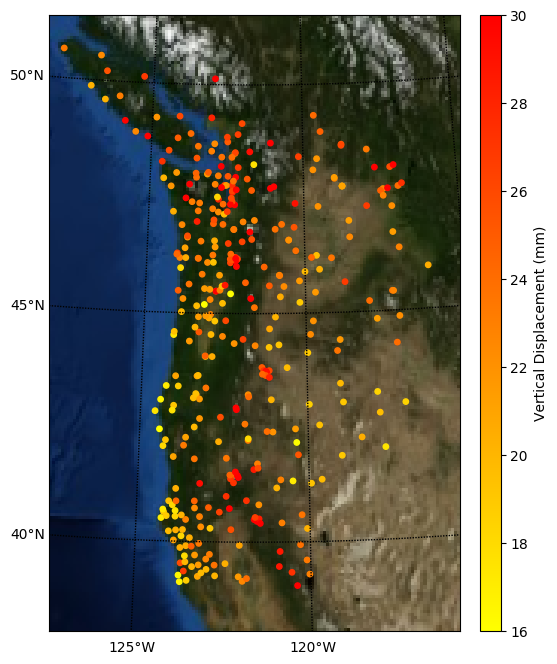}}\hfill
	\subfloat[Horizontal displacement]{\label{fig:Cascadia_h}\includegraphics[height=0.35\textheight]{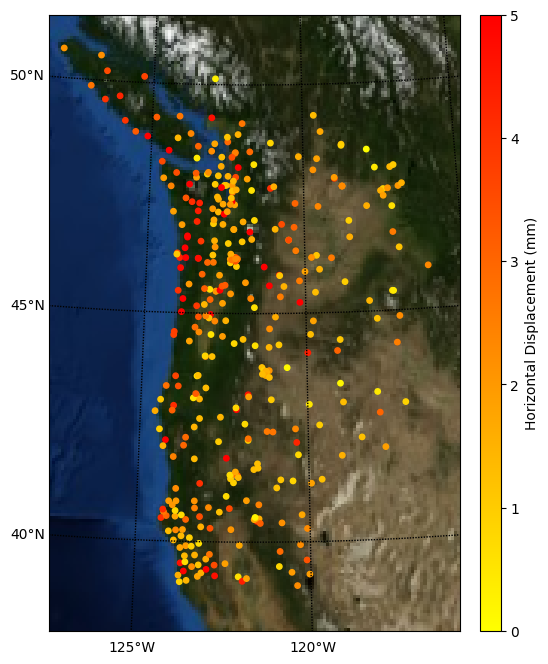}}\\
	\subfloat[Vertical-Horizontal Ratio]{\label{fig:Cascadia_ratio}\includegraphics[height=0.35\textheight]{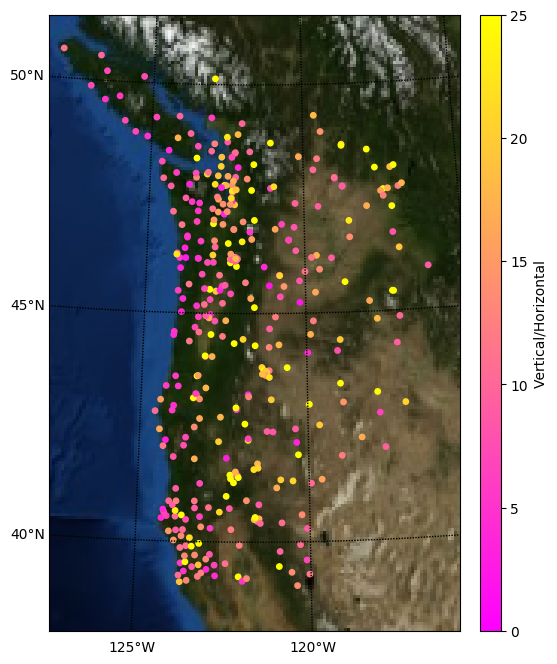}}\hfill
	\subfloat[Tilt]{\label{fig:Cascadia_tilt}\includegraphics[height=0.35\textheight]{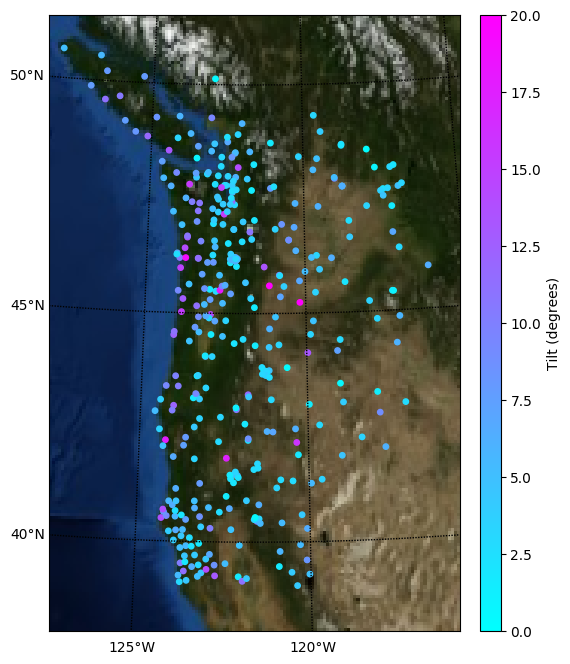}}
	\caption{\textbf{Measures of surface deformation in Cascadia subduction zone.} \textbf{a,} Net vertical displacement and \textbf{b,} net horizontal displacement computed from GPS measurements, \textbf{c,} their ratio, and \textbf{d,} hodogram tilt (in degrees) from the vertical. All color scales have been truncated to expose the trends.}
	\label{fig:Cascadia}
\end{figure}

We generate hodograms for all the GPS measurements at sites in the Cascadia subduction zone and in Alaska and thereafter compute the vertical displacement, horizontal displacement, their ratio, and the hodogram tilt.
These attributes for Alaska and Cascadia are shown in Figures~\ref{fig:Alaska} and \ref{fig:Cascadia}, respectively.
Note that in both cases, the horizontal displacement decreases monotonically from the margin landwards; while the vertical displacement increases as one moves landwards from the margin, attains a maximum, and decreases thereafter.
The above deformation trends closely correspond to those seen in numerical simulation of deformation seen in Section~\ref{Buckling} and analyzed above in Figure~\ref{fig:GPSandHodo}.

The belt of maximum vertical displacements along the Cascadia margin has a close correspondence to the tremor maps generated by \citeA{Wech2009,Wells2017}.
Similarly, the maximum vertical displacements in Alaska encompass the tremor activity mapped by \citeA{Ohta2006} and \citeA{Peterson2009} (in addition to showing locations where additional tremor activity could be expected).

The tilt of the hodogram major-axis (Figures~\ref{fig:Alaska_tilt} and \ref{fig:Cascadia_tilt}) shows a trend that is consistent with what one would expect to observe for buckling (Figures~\ref{fig:A_Hodo}, \ref{fig:B_Hodo}, \ref{fig:C_Hodo}, and \ref{fig:D_Hodo}).
The trend is especially prominent for Alaska, where the tilts show values as high as 30$^\circ$ close to the coastline and systematically decrease as one moves inland, attaining values close to 0$^\circ$ over the tremor zones.

With regards to vertical GPS measurements, we observe that
\begin{itemize}
	\item their amplitudes can be large and in many cases an order of magnitude larger than horizontal displacements,
	\item there is a close correspondence between sudden changes in horizontal displacements (horizontal GPS reversals) and rapid vertical GPS measurements on numerous occasions, and
	\item vertical displacement patterns (Figure~\ref{fig:Alaska} and \ref{fig:Cascadia}) show close spatial correspondence with spatial tremor patterns in Cascadia and Alaska.
\end{itemize}
Given the above observations, we conclude that the observed vertical displacements contain significant imprints of tectonic deformation from buckling and collapse.

\section{Episodic Buckling and Collapse} \label{EBC}
Based on the novel observations presented above, combined with the numerical modeling of deformation, and spatio-temporal analysis of GPS in Cascadia and Alaska, we introduce an Episodic Buckling and Collapse model of the subduction zone, whereby periodic seismic activity (tectonic tremor) and geodetic changes, result from the episodic buckling of the overriding continental crust and its rapid collapse on the subducting oceanic slab.

However, it is not sufficient that the proposed model fits only the new observations and analysis presented above; the model should also be able to clearly explain the numerous observations and findings already published.
Until now, all these observations have been explained in the light of the Slow Slip hypothesis.

\subsection{Published Observations and Findings} \label{existingPubs}
Table~\ref{table:Comparison} summarizes various geodetic observations, seismological studies, imaging research, and geologic findings, all of which should provide constraints for any model of the subduction zone.
Therefore, below we list all these scientific findings and also discuss how they fit into the Slow Slip hypothesis.
Thereafter, in section~\ref{EBCModel}, we use these observations in conjunction with the observations presented above to develop the Episodic Buckling and Collapse model and explain how other scientific findings are reasonably explained by it.

\begin{table}
	\centering
	\begin{tabular}{ p{6cm} p{9.0cm}}
		\toprule\toprule
		\textbf{Observations} 						& \textbf{References} \\ \midrule\midrule
		GPS Horizontal								& \citeA{Hirose1999,Dragert2001}\\ \midrule
		GPS Vertical   								& \citeA{Douglas2005,Miyazaki2006,Heki2008,Fu2013,Liu2015,Behura2018}\\ \midrule
		Tiltmeter recordings						& \citeA{Obara2004,Hirose2005,Hirose2010}\\ \midrule
		Presence of fluids in LVZ  					& \citeA{Eberhart-Phillips2006,Matsubara2009,Audet2009,Kim2009,Bell2010,Hansen2012,Toya2017}\\ \midrule
		Large fluid pressure in LVZ   				& \citeA{Audet2009,Kim2009,Toya2017}\\ \midrule
		Low effective stress   						& \citeA{Rubinstein2007,Bell2010}\\ \midrule
		Episodic fluid drainage   					& \citeA{Nakajima2018}\\ \midrule
		LFEs and VLFEs   							& \citeA{Liu2015,Frank2019}\\ \midrule
		Thick LVZ   								& \citeA{Hansen2012,Toya2017,Audet2018}\\ \midrule
		LVZ Geometry and their up-dip \& down-dip extents	& \citeA{Matsubara2009,Hansen2012,Toya2017,Audet2018}\\ \midrule
		Occurrence of tremors   					& \citeA{Obara2002}\\ \midrule
		Tremor source mechanism   					& \citeA{Shelly2006,Wech2007,Bostock2012,Ohta2019}\\ \midrule
		Spatial extent of tremors   				& \citeA{Matsubara2009,Wech2009,Kao2009,Audet2010,Audet2018}\\ \midrule
		Tremor migration patterns   				& \citeA{Shelly2007,Wech2009,Kao2009,Ghosh2010,Boyarko2010,Obara2011b,Obara2012}\\ \midrule
		Absence of tremors on old crusts  			& \citeA{Schwartz2007}\\ \midrule
		Variable tremor and slow slip periodicity  	& \citeA{Wallace2010}\\ \midrule
		Tremors located down-dip of LVZ  			& \citeA{Peterson2009,Audet2018}\\ \midrule
		Crustal seismicity   						& \citeA{Nicholson2005,Shelly2006,Bostock2012}\\ \midrule
		Mantle helium correlation with tremor location	& \citeA{Umeda2007,Sano2014}\\ \midrule
		Paleo-uplift and subsidence   				& \citeA{Dragert1994,Sherrod2001,Leonard2004,Shennan2006}\\ \midrule
		\bottomrule
	\end{tabular}
	\caption{List of observations and results used in and explained by constructing the Episodic Buckling and Collapse model of subduction zones.}
	\label{table:Comparison}
\end{table}

\subsubsection{Geodetic Observations}
In addition to the reversals in horizontal GPS recordings, similar and more prominent reversals are observed on the vertical GPS component \cite{Douglas2005,Miyazaki2006,Heki2008,Behura2018}.
Magnitude of the vertical displacements cannot be satisfactorily explained by the Slow-Slip hypothesis as it assumes only relative sliding between the subducting slab and the overriding plate.

Tiltmeter recordings \cite{Obara2004,Hirose2005,Hirose2010} show significant bulging of the surface prior to slow slip and subsequent contraction coinciding with slow slip.
Although temporal changes in tiltmeter recordings can be reasonably explained by the Slow-Slip hypothesis, accounting for the spatial changes through slow slip is more challenging.

\subsubsection{Fluids and the Low-Velocity Zone}
Numerous studies \cite{Eberhart-Phillips2006,Matsubara2009,Audet2009,Bell2010,Toya2017} clearly demonstrate the presence of fluids at the plate interface characterized by a seismic Low-Velocity Zone (LVZ).
It is widely believed that slab dehydration generates aqueous fluids which then travel upward because of buoyancy forces and accumulate at the plate interface and mantle wedge.
Seismologists believe that these fluids lubricate the plate interface, thereby aiding slow slip and aseismic slip.

In Cascadia, evidence of fluids come from the work of \citeA{Audet2009} who employ teleseismic data to show the presence of a zone with anomalously high Poisson's ratio extending from the margin all the way to the corner of the mantle wedge.
Presence of fluids in the tremor region in Shikoku is evident from the tomographically-derived low velocities by \citeA{Shelly2006} and \citeA{Matsubara2009}.

Other studies show that the plate interface is overpressured \cite{Audet2009,Toya2017}.
\citeA{Rubinstein2007,Bell2010} find extremely low effective normal stresses in subduction zones.
Excepting buoyancy recharging the plate boundary with hydrous magmatic fluids, the Slow-Slip model provides little explanation of the cause of overpressure and their periodic nature.

Recent findings by \citeA{Nakajima2018} shed new light on the movement of fluids at the plate boundary.
They analyze seismic data spanning more than a decade over Japan and demonstrate that ``seismicity rates and seismic attenuation above the megathrust of the Philippine Sea slab change cyclically in response to accelerated slow slip.''
They interpret these findings to represent ``intensive drainage during slow slip events that repeat at intervals of approximately one year and subsequent migration of fluids into the permeable overlying plate.''
Although \citeA{Nakajima2018} provide an explanation of these observation in the context of the Slow-Slip hypothesis, it is unclear what forces drive the fluids in and out of the plate boundary.

The spatial extent and geometry of the LVZ are clear from the work of \citeA{Hansen2012,Toya2017,Audet2018}.
\citeA{Toya2017,Audet2018} report a thick LVZ with thicknesses averaging a few kilometers in the Cascadia Subduction Zone.
All of them also report the thickening of the LVZ with increasing depth.
It is unclear how such a thick ductile zone could be generating tremor.
\citeA{Audet2018} also note that the LVZ does not extend into the locked zone; and on the down-dip side, it truncates at the mantle wedge.
They conclude that the nature of the LVZ remains ambiguous and provide a couple of hypothesis explaining the increasing thickness of the LVZ with depth.
These hypothesis, however, do not provide a definitive explanation of the periodic nature of slow slip.

\subsubsection{Tremor}
Since the first reporting by \citeA{Obara2002}, tremor in subduction zones has been widely observed all over the world.
Several researchers have reported that tremor has a dominant thrust-type focal mechanism \cite{Shelly2006,Wech2007,Bostock2012}, thereby providing a significant boost to the proponents of the Slow-Slip hypothesis.
According to the Slow-slip hypothesis, as the subducting slab slides underneath the continental crust during slow slip, it generates tremor with predominant thrust-type focal mechanism.


Tectonic tremors are usually located in a narrow spatial interval oriented in a strike-parallel direction \cite{Wech2009,Kao2009,Obara2010,Audet2010,Audet2018}.
The down-dip boundary is close to the mantle wedge, while the up-dip boundary extends a few kilometers from the mantle wedge.
In southwestern Japan, \citeA{Matsubara2009} observe that, ``These tremors occur at the landward
edge of the high-$V_P/V_S$ zone only beneath the southern Kii peninsula. The common point of the tremors for these four regions is that the tremors are distributed in places where the Philippine Sea plate first contacts with the serpentinized wedge mantle of the Eurasian plate.''
In the light of the Slow-Slip model, multiple explanations of their depth extent have been proposed, all of them revolving around variations in slip properties of the plate boundary due to temperature and pressure changes.

Multiple studies \cite{Peterson2009,Audet2018} image the tremor swath to the down-dip side of the LVZ.
\citeA{Audet2018} interpret these observations as reflective of transitions in plate coupling and slip modes along the dip.
If such transitions are indeed present, the processes that result in such changes along the plate boundary are open to question.

Tremors exhibit peculiar migration characteristics.
\citeA{Wech2009,Obara2011b} observe up-dip and radial tremor migration.
\citeA{Obara2010,Obara2012} show a bimodal distribution of tremors in the Nankai subduction zone, with tremors from the along-strike migration concentrated on the up-dip side, while tremors from up-dip migration distributed over the entire tremor zone.
Other studies \cite{Houston2011,Obara2012} report rapid reverse tremor migration where tremors migrate in the opposite direction of along-strike migration at much faster speeds.
It is unclear from the Slow-Slip hypothesis as to what physical phenomena might result in such migration patterns.

\citeA{Schwartz2007} find no evidence of slow slip and tremors in northeast Japan which has a thick old crust, while younger and thinner crusts in the Nankai subduction zone exhibit an array of slow slip events with varying periodicity.
\citeA{Wallace2010} report an interesting correlation between temporal characteristics of slow slip events and their depth of occurrence in the Hikurangi subduction margin of New Zealand.
They note that the longest duration, and largest slow slip events occur at large depths, while the shortest duration, smallest, and most frequent slow slip events are usually shallow.
Although the degree of plate coupling \cite{Wallace2010} can explain some of these observations, it is unclear how plate coupling can explain the variable periodicity and duration of the slow slip events.

\subsubsection{Crustal Seismicity}
Significant crustal seismicity is observed in Cascadia \cite{Nicholson2005,Kao2005a,Bostock2012} and Nankai \cite{Shelly2006} subduction zones.
A majority of the reported crustal seismicity is located at shallow depths and a few kilometers above the tremor zone and further landward.
The Slow-Slip hypothesis does not provide a satisfactory explanation either of the origin of such seismicity or for the spatial correspondence between shallow crustal seismicity and deep tremor.

\subsubsection{Mantle Helium}
\citeA{Sano2014} report interesting findings and suggest the existence of fluid pathways from the mantle to the trench in the Nankai subduction zone.
They note, ``a sharp increase in mantle-derived helium in bottom seawater near the rupture zone 1 month after the earthquake. The timing and location indicate that fluids were released from the mantle on the seafloor along the plate interface. The movement of the fluids was rapid, with a velocity of $\approx$4 km per day and an uncertainty factor of four. This rate is much faster than what would be expected from pressure-gradient propagation, suggesting that over-pressurized fluid is discharged along the plate interface.''
It is debatable as to what forces mantle fluids to squirt out in the vicinity of the rupture zone during megathrust earthquakes.

Furthermore, \citeA{Umeda2007} observe a close spatial correspondence between mantle helium and tremors.
They report a high flux of mantle helium over regions experiencing tremors and a low flux in areas adjacent to those lacking tremors.
Reconciling these observations with slow slip had proved to be challenging.

\subsubsection{Paleo-Uplift and Subsidence}
Evidence of large-scale and periodic continental deformation can be found in geologic records.
\citeA{Sherrod2001} find evidence of abrupt sea level changes and rapid submergence in Puget Sound, Washington State.
They estimate a maximum subsidence of approximately 3~m.
\citeA{Leonard2004} report a maximum subsidence of 2~m during the 1700 great Cascadia earthquake.
In Alaska, \citeA{Hamilton2005a,Hamilton2005b,Shennan2006} report rapid subsidence measuring 2~m.
It is unclear from the Slow-Slip model as to how the crust can experience an uplift in excess of 2~m over a period of 500~years.

\subsection{The Model} \label{EBCModel}
The Slow-Slip hypothesis depicts a plate interface that is frictionally locked at shallow depths and transitions into a slow-slip zone down-dip.
Below this transition zone, geoscientists believe that the subducting slab slides continuously at a steady rate consistent with plate motion.
\emph{The key assumption in these models is that the overriding continental plate is in physical contact with the subducting oceanic slab all along the plate interface.}

The Episodic Buckling and Collapse model, on the other hand, is based on the hypothesis of a buckling overriding plate that detaches itself down-dip from the subducting slab, while being in contact in the locked seismogenic zone.
According to this model, the observed low-velocity zone (LVZ) is neither a part of the continental crust nor the subducting slab.
Instead, it is a fluid-filled chamber created between the two plates because of the buckling of the overriding continental plate.
An interplay of plate deformation, pressure differentials, and pressure release control the fluid flow in and out of this chamber and also generate seismicity in the form of tectonic tremor, low-frequency, and very-low-frequency energy releases.

Below, we describe the various temporal phases of the short-term buckling and collapse process within each cycle and the multiple physical phenomena occurring within each of the phases.
Later we introduce long-term buckling and collapse cycles that are related to megathrust cycles.
Section~\ref{MegathrustEBC} dwelves further into this hypothesis and provides a potential link between Episodic Buckling and Collapse and megathrust earthquakes.

\subsubsection{Phase \textbf{T}$\bm{_0}$}
Because only the seaward edge of the plate interface (accretionary wedge and seismogenic zone) is `locked' while the rest of the interface can slide, the overriding plate will buckle under the forces of the subduction process.
Given the slowly developing subduction processes, the system will exhibit Euler's fundamental model of buckling -- with the locked portion of the continental plate acting as one fixed end and the thick continental crust further inland serving as the other fixed end of the buckling system.
Figure~\ref{fig:Buckling} shows a schematic of the buckling and collapse process occurring in subduction zones.
Phase \textbf{T}$\bm{_0}$ corresponds to a state within the buckling cycle where the tectonic stresses on the overriding continental plate are minimal (phase \textbf{T}$\bm{_0}$, Figure~\ref{fig:Buckling}).
A magmatic-fluid-filled chamber exists between the overriding plate and the subducting slab.

\subsubsection{Phase \textbf{T}$\bm{_1}$}
As the oceanic slab subducts, compressive stresses build up within the overriding plate, thereby pushing it upward and landward (phase \textbf{T}$\bm{_1}$, Figure~\ref{fig:Buckling}).
The overriding plate starts buckling further to accommodate the additional strain, wherein the deep continental crust overlying the transition zone and the mantle wedge buckles away from the subducting slab and possibly the mantle.

\textbf{Fluid Flow} -- 
The above deformation enlarges the size of the fluid-filled chamber and drives down the pore-pressure inside it, which in turn results in upwelling of magmatic fluids from the wedge region towards the chamber (Figure~\ref{fig:MagmaFlow_T1}).
This process is slow and occurs for majority of the cycle.
For example, in Cascadia, phase \textbf{T}$\bm{_1}$ continues for majority of the 14 months.
Because this phase evolves slowly, pressure equilibrium is maintained throughout the phase as progressive buckling is accompanied by steady fluid upwelling.

\textbf{Low Effective Stress} -- 
We expect the effective stress of the system to be close to zero and any small stress perturbations may lead to escape of fluids through faults, fractures, fissures (and potentially magma vents), and also result in minor collapse of the overriding plate thereby generating tremor.
Evidence of low effective normal stress comes from observations that tremors may not only be triggered by earthquakes \cite{Brodsky2007,Miyazawa2008,Rubinstein2007,Peng2008} but also, more interestingly, by tides \cite{Shelly2007,Rubinstein2008,Hawthorne2010}.

Surface bulging due to buckling is consistent with the tiltmeter measurements (phases \textbf{T}$\bm{_2}$, \textbf{T}$\bm{_3}$, and \textbf{T}$\bm{_4}$, Figure~\ref{fig:Buckling}) as reported by \citeA{Hirose2005,Obara2004} who observe that the surface is dome shaped during tremor episodes.
It would be interesting to study and quantify the temporal evolution in spatial patterns of tiltmeter measurements.

\subsubsection{Phase \textbf{T}$\bm{_2}$}
Progressive buckling will result in continual opening of faults and fractures, with the openings starting at shallow depths and progressing downwards.
At a certain critical state, right before the fracture and fault openings reach the fluid-filled chamber, buckling exhibits the maximal horizontal and vertical displacements of the overriding plate (phase \textbf{T}$\bm{_2}$, Figure~\ref{fig:Buckling}) within each cycle.

Phase \textbf{T}$\bm{_2}$ also corresponds to the maximal extensional stress on the top of the overriding plate and the maximal volume of the fluid chamber within each cycle.
The structure of the fluid chamber would be similar to what has been observed by \citeA{Hansen2012,Toya2017,Audet2018} -- thickening of the LVZ with increasing depth.
Our model suggests that the LVZ extends into the continental Moho and truncates to the landward-side of the mantle wedge.
The weak continental Moho reflectivity observed in the Cascadia subduction zone by \citeA{Haney2016} is evidence of the LVZ extending landward into the continental Moho.
Detailed imaging studies are needed to establish the precise landward-extent of this fluid chamber.

The time between Phases \textbf{T}$\bm{_0}$ and \textbf{T}$\bm{_2}$ corresponds to gradual buckling and slow upwelling of fluids.
Such gradual deformations and steady fluid flow do not emanate any seismic energy in the vicinity of the plate boundary.
However, the continual buckling and bulging of the overriding continental plate result in opening of strike-parallel and transverse faults resulting in significant crustal seismicity as observed by \citeA{Nicholson2005,Shelly2006,Bostock2012}.
The shallow crust is expected to house a majority of this seismicity because it experiences the maximum strain.

\subsubsection{Phase \textbf{T}$\bm{_3}$}
\begin{figure}
	\center
	\subfloat[Phase \textbf{T}$\bm{_1}$]{\label{fig:MagmaFlow_T1}
		\begin{overpic}[width=\textwidth,trim=20mm 70mm 30mm 0,clip=true]{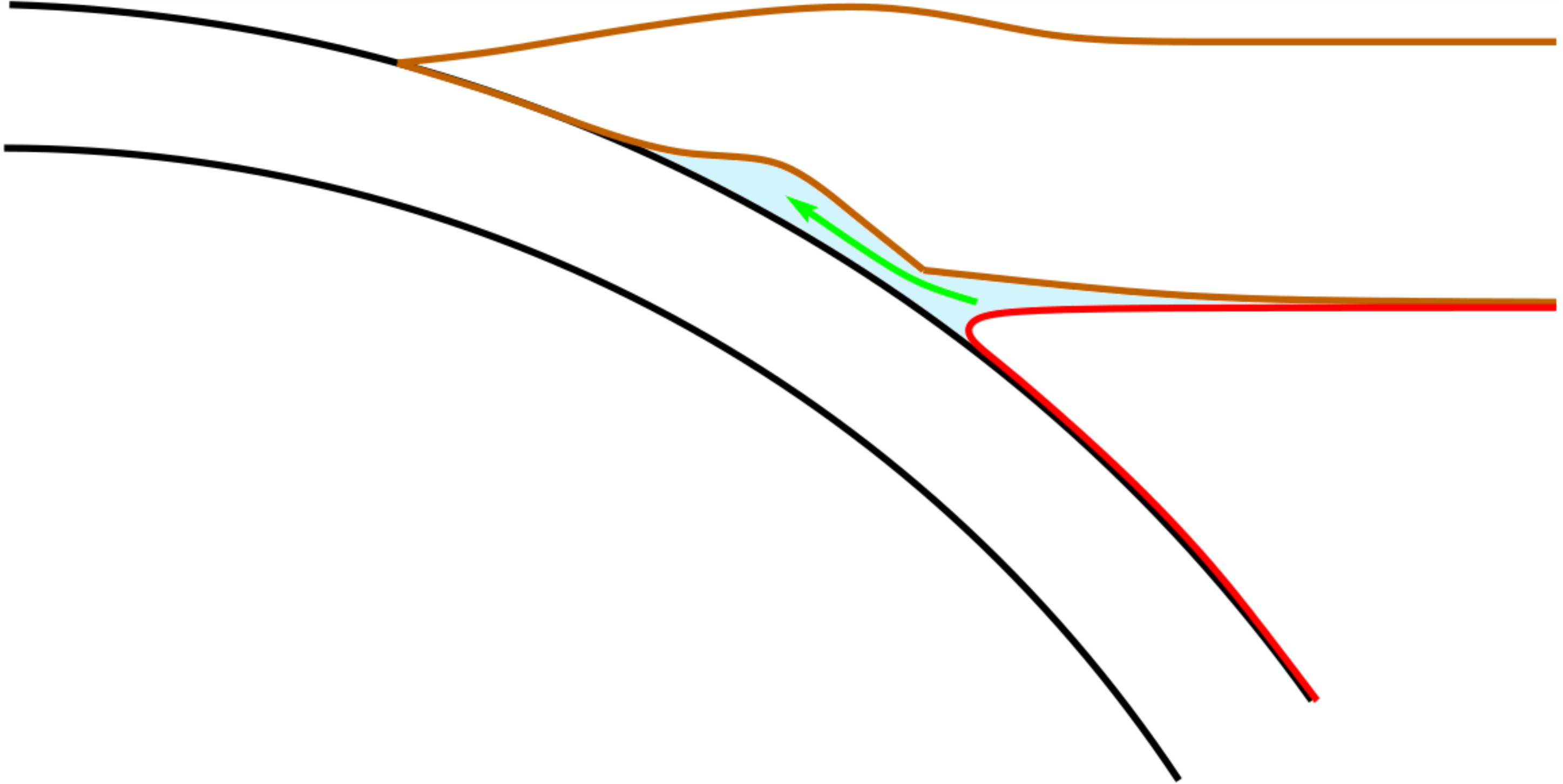}
		\thicklines
		\put(5,20){\color{black}\scalebox{2}{\rotatebox[origin=c]{-11}{$\rightarrow$}}}
		\put(50,3){\color{black}\scalebox{2}{\rotatebox[origin=c]{-35}{$\rightarrow$}}}
		\put(55,20){\color{red}\scalebox{1}{$\rightarrow$}}
		\put(53,24){\color{blue}\scalebox{1}{$\uparrow$}}
		\put(23,21){\color{black}\scalebox{2}{\rotatebox[origin=c]{70}{$\Big\updownarrow$}}}
		\put(23,19){\color{black}\scalebox{1}{\rotatebox[origin=c]{-20}{\myfont locked}}}
		\put(85,17){\textbf{\myfont T$\bm{_1}$}}
		\end{overpic}}\\[2mm]
	\subfloat[Phase \textbf{T}$\bm{_3}$]{\label{fig:MagmaFlow_T3}
		\begin{overpic}[width=\textwidth,trim=20mm 70mm 30mm 0,clip=true]{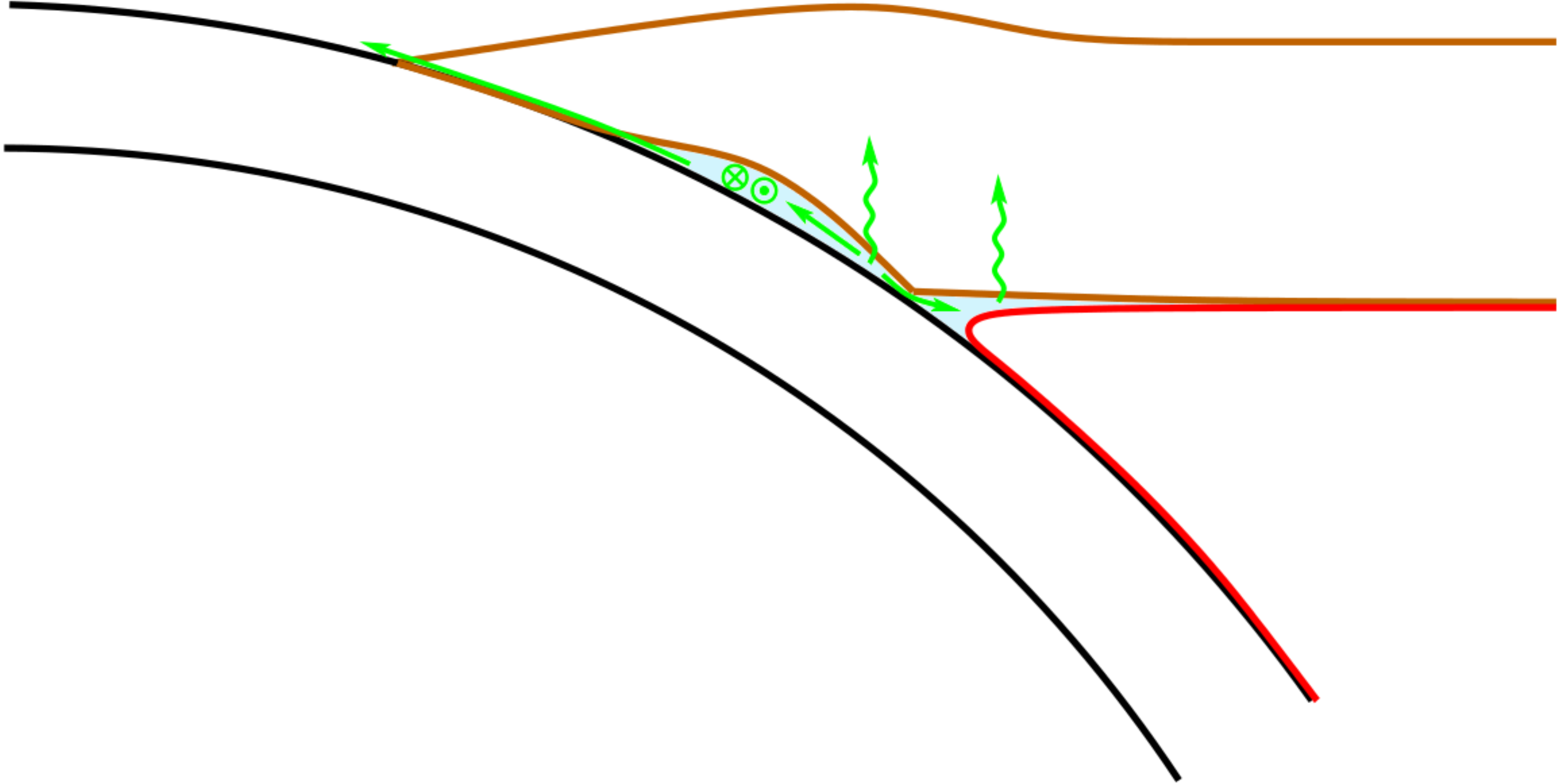}
		\thicklines
		\put(4,21){\color{black}\scalebox{1}{\rotatebox[origin=c]{171}{$\rightarrow$}}}
		\put(45,7){\color{black}\scalebox{1}{\rotatebox[origin=c]{149}{$\rightarrow$}}}
		\put(45,24){\color{red}\scalebox{2}{$\leftarrow$}}
		\put(51,20){\color{blue}\scalebox{2}{$\downarrow$}}
		\put(23,21){\color{black}\scalebox{2}{\rotatebox[origin=c]{70}{$\Big\updownarrow$}}}
		\put(23,19){\color{black}\scalebox{1}{\rotatebox[origin=c]{-20}{\myfont ?locked?}}}
		\put(85,17){\textbf{\myfont T$\bm{_3}$}}
		\end{overpic}}\\[1mm]
	\vspace{-2mm}
	\caption{\textbf{Schematic illustration of magmatic fluid flow during each Episodic Buckling and Collapse cycle.} Two-dimensional cross-sections for phases (a) \textbf{T}$\bm{_1}$ and (b) \textbf{T}$\bm{_3}$ are shown. Green arrows represent the flow direction of magmatic fluids. The sinuous green lines fluid flow into the overlying crust.}
	\label{fig:MagmaFlow}
\end{figure}
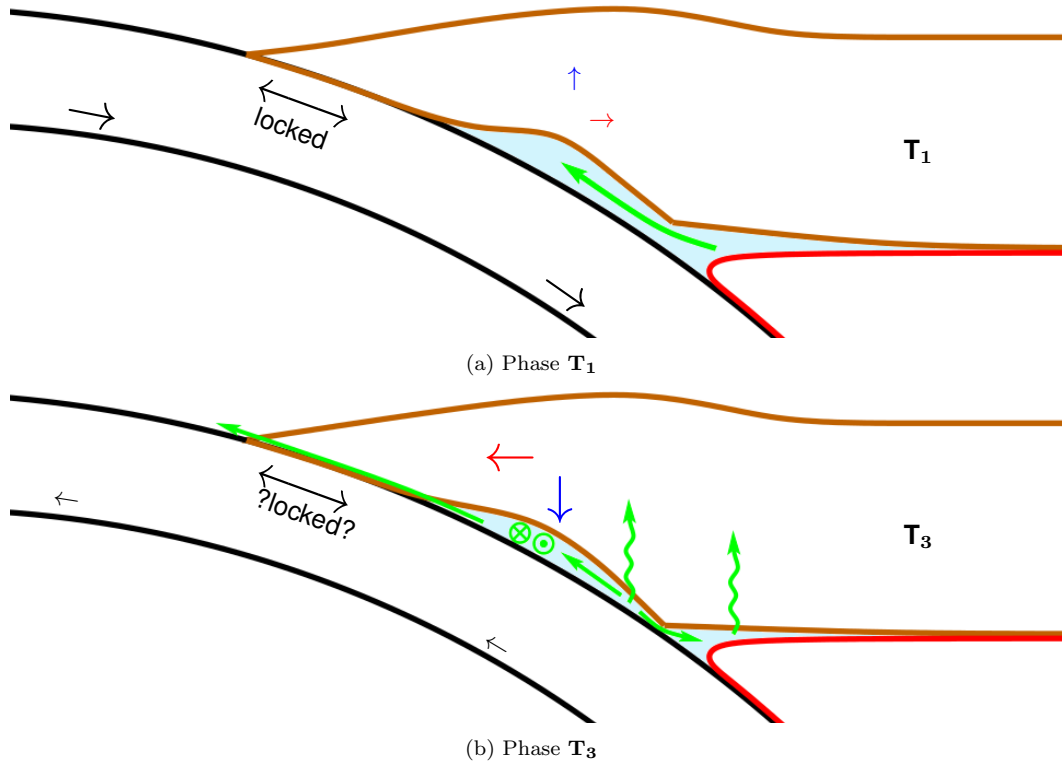

\textbf{Fluid Chamber Collapse} -- 
As soon as the fault and fracture openings reach the fluid-filled chamber, the magmatic fluid escapes into the overriding plate (most likely accompanied by phase change from liquid to gaseous) and consequently drops the pressure inside the chamber dramatically (Figure~\ref{fig:MagmaFlow_T3}).

As a result, the chamber starts collapsing as illustrated in phase \textbf{T}$\bm{_3}$ of Figure~\ref{fig:Buckling}.
The rapid reversal observed in horizontal GPS measurements is a result of the collapse-related seaward horizontal displacement and not from so-called slow slip.
As shown below and as expected, changes in vertical displacement are even more substantial.

\citeA{Wells2017} demonstrate substantial evidence of regional faults extending to the plate interface.
The distribution of mantle helium in eastern Kyushu by \citeA{Umeda2007} is consistent with the above picture.
\citeA{Umeda2007} observe a close correspondence of mantle helium (in hot springs) with the occurrence of tremor -- the flux of mantle helium is low in areas lacking tremors, while it is high above regions experiencing tremors.

\textbf{Fluid Flow} -- 
The rapid collapse of the continental plate will dramatically increase the fluid pressure inside the chamber, which in turn will push the fluid up-dip, down-dip, and along-strike (phases \textbf{T}$\bm{_3}$, and \textbf{T}$\bm{_4}$, Figures~\ref{fig:Buckling} and \ref{fig:MagmaFlow}).

Also, there is a distinct possibility that the high fluid pressure fluids breaks flow barriers within conduits and asperities housed in the locked zone and the accretionary prism, leading to the up-dip escape of some magmatic fluids along the locked zone through the accretionary prism (Figure~\ref{fig:MagmaFlow_T3}).
The collapsing continental plate will also push fluids along-strike at the plate boundary as shown below in phase \textbf{T}$\bm{_4}$.

\textbf{VLFEs} -- 
We hypothesize that the so-called shallow very-low-frequency earthquakes (VLFEs) observed in accretionary prisms result from the rapid flow of magmatic-fluid brought about by the collapsing continental crust.

Multiple researchers have reported the close spatial and temporal correspondence of shallow very-low-frequency earthquakes (VLFEs) in the accretionary prism with deep tremor and short-term slow slip events.
\citeA{Obara2005} report shallow VLFEs on the up-dip side of the locked zone in the Nankai trough.
Because the accretionary prism contains out-of-sequence thrusts and fault splays, \citeA{Obara2005} speculate that these fault planes might provide pathways for fluid flow from the subducting slab.
More recently, the work of \citeA{Liu2015,Nakano2018} shows the close temporal association between shallow VLFEs in the accretionary prism with deep short-term slow slip events.
\citeA{Liu2015} provide clear evidence of the occurrence of VLFEs predominantly at the onset of short-term slow slip.
They also show that these VLFEs have thrust-type focal mechanism.
Note that \citeA{Liu2015} assume a moment-tensor source mechanism in their inversions, and not single point forces.
However, we believe that the one should use single point forces as the source mechanisms for VLFEs, which would yield the direction and intensity of fluid flow.

We do not expect any seismicity at the plate boundary (due to plate motion or fluid flow) during the buckling phase (phases \textbf{T}$\bm{_0}$, and \textbf{T}$\bm{_1}$, Figure~\ref{fig:Buckling}) but expect different forms of energy release (at multiple locations on the plate boundary) during the collapse phases (phases \textbf{T}$\bm{_3}$, and \textbf{T}$\bm{_4}$, Figure~\ref{fig:Buckling}) arising from plate striking as well as fluid flow.

\textbf{Other Explanations for Chamber Collapse} -- 
The locked zone experiences substantial stress because of the buckling continental plate.
Another possible scenario for the overriding plate collapse could be the minor and temporary decoupling of the locked zone when frictional forces in the locked zone are exceeded.
Focal mechanisms of such seismic activity should be close to thrust-type.
However, the lack of significant conventional seismicity (high frequency) in the locked zone prior to tremors is a strike against this possibility.
Any future discovery of locked-zone conventional seismicity immediately preceding tremor activity will add substantial credibility to this potential scenario.

It is also possible that a combination of the above two processes -- fluid flow and locked-zone decoupling, might be occurring.
Future research efforts on understanding the dynamic processes at locked zone and the accretionary prism will shed more light on the dominant mechanism.

\subsubsection{Phase \textbf{T}$\bm{_4}$}
\textbf{Tectonic Tremor Origin} -- 
The rapidly collapsing overriding plate strikes the subducting oceanic slab, thereby generating tectonic tremor (phase \textbf{T}$\bm{_4}$ of Figure~\ref{fig:Buckling} and Figure~\ref{fig:MagmaFlow_T4}).
Tremor source mechanisms at subduction zones should therefore be predominantly of the Compensated Linear Vector Dipole (CLVD) type, with a possible minor thrusting component arising from the relative plate motion.
Researchers have, however, observed a dominant thrust-type focal mechanism for tremor \cite{Shelly2006,Wech2007,Bostock2012}.
That being said, in the absence of full-azimuth and wide-angle sampling of a focal sphere, one might mistake a CLVD mechanism as a thrust-type mechanism.

Fluid flow can further complicate the estimation of a source mechanism.
Fluid motion, by itself, has a source mechanism of a single point force (and not a moment tensor).
Moreover, if the fluid is viscous, the fluid drag against the plate walls will result in a thrust-type and/or normal-fault-type focal mechanism.
Because LFEs and tremors usually accompany each other \cite{Shelly2006,Brown2013}, it is quite likely that all these source mechanisms are superimposed on top of one another, making the inversion and interpretation of tremor source mechanisms challenging.

The atypical lower-boundary geometry of the buckled continental plate explains why tremors truncate at the continental Moho (phase \textbf{T}$\bm{_4}$, Figures~\ref{fig:Buckling} and \ref{fig:MagmaFlow_T4}) and are observed lying within a narrow band up-dip along the plate interface \cite[phase \textbf{T}$\bm{_4}$, Figures~\ref{fig:Buckling} and \ref{fig:MagmaFlow_T4},][]{Wech2009,Peterson2009,Audet2018}.
\citeA{Audet2010} observe that ``the peak occurrence of tremors roughly coincides with the intersection of the plate interface with the overlying continental crust--mantle boundary''.

\textbf{Fluid Flow} -- 
As the overriding continental crust collapses with the lower edge hitting the subducting slab first, some of the fluids are pushed landward along the continental Moho, while most of the fluids are pushed up-dip and along-strike (Figure~\ref{fig:Buckling} and \ref{fig:MagmaFlow_T4}).
It is likely that as the lower edge hits the subducting slab, it cuts off hydraulic communication between the up-dip fluid chamber and the down-dip mantle wedge, thereby trapping fluid in the chamber.
As described above, the collapse also increases the pore-pressure in the chamber, without which the up-dip rate of collapse (parameter that controls tremor migration rate) would be larger than the ones observed by \citeA{Wech2009} and \citeA{Obara2011b} in Cascadia and Japan, respectively.
In the latter part of phase \textbf{T}$\bm{_4}$, the lagging end of the high-pressure fluid pocket collapses (creating tremors), thereby pushing the fluid pocket up-dip and parallel to the strike along the plate boundary.

Similar to shallow VLFEs, we hypothesize that deep low-frequency earthquakes (LFEs and VLFEs), observed by many researchers \cite{Ito2007,Ito2009,Matsuzawa2009,Obara2011a,Frank2019}, correspond to the rapid sloshing of magmatic fluids brought about by the hastened collapse of the overriding plate.
\citeA{Frank2019} demonstrate the remarkable spatio-temporal correlation with slow slip events.
Also interestingly, they show that the magnitude of the LFE is maximum in the vicinity of the mantle wedge -- exactly as predicted by the Episodic Buckling and Collapse model.
The up-dip location of deep VLFEs with respect to that of tremor indicates that most of the magmatic fluid is pushed up-dip in phases 3 and 4 (Figures~\ref{fig:MagmaFlow_T3} and \ref{fig:MagmaFlow_T4}).

\subsubsection{Phases \textbf{T}$\bm{_5}$ and \textbf{T}$\bm{_6}$}
\begin{landscape}
\begin{figure}
	\center
	\subfloat{\label{fig:T0}
	\begin{overpic}[width=0.75\textwidth,trim=20mm 70mm 30mm 0,clip=true]{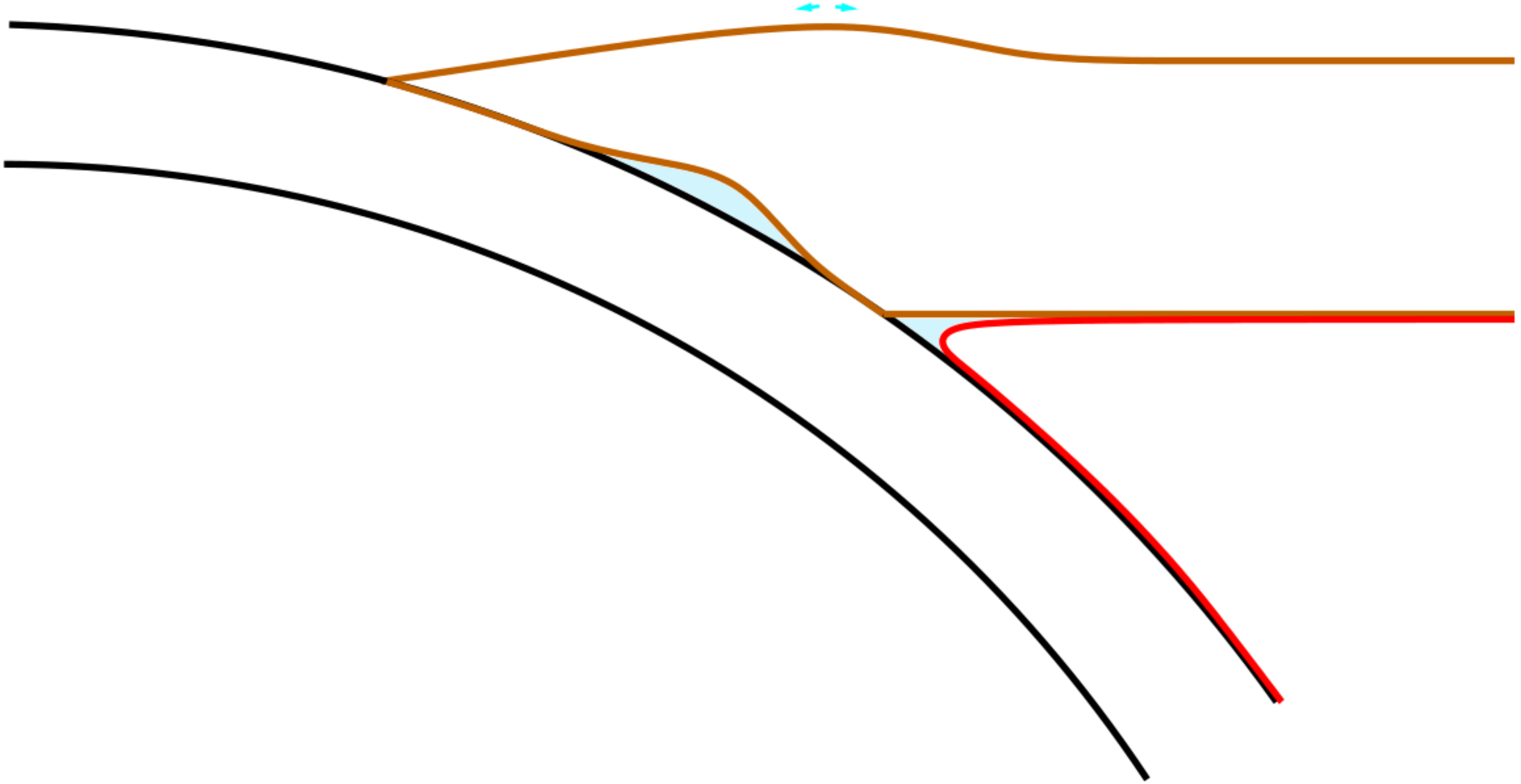}
		\thicklines
		\put(5,20){\color{black}\scalebox{2}{\rotatebox[origin=c]{-11}{$\rightarrow$}}}
		\put(50,3){\color{black}\scalebox{2}{\rotatebox[origin=c]{-35}{$\rightarrow$}}}
		\put(55,20){\color{red}\scalebox{1}{$\rightarrow$}}
		\put(53,24){\color{blue}\scalebox{1}{$\uparrow$}}
		\put(22,21){\color{black}\scalebox{2}{\rotatebox[origin=c]{72}{$\Big\updownarrow$}}}
		\put(22,19){\color{black}\scalebox{1}{\rotatebox[origin=c]{-18}{\myfont locked}}}
		\put(25,35){\myfont A}
		\put(40,35){\myfont B}
		\put(55,35){\myfont C}
		\put(70,35){\myfont D}
		\put(85,17){\textbf{\myfont T$\bm{_0}$}}
	\end{overpic}}\hfill
	\subfloat{\label{fig:T3}
		\begin{overpic}[width=0.75\textwidth,trim=20mm 70mm 30mm 0,clip=true]{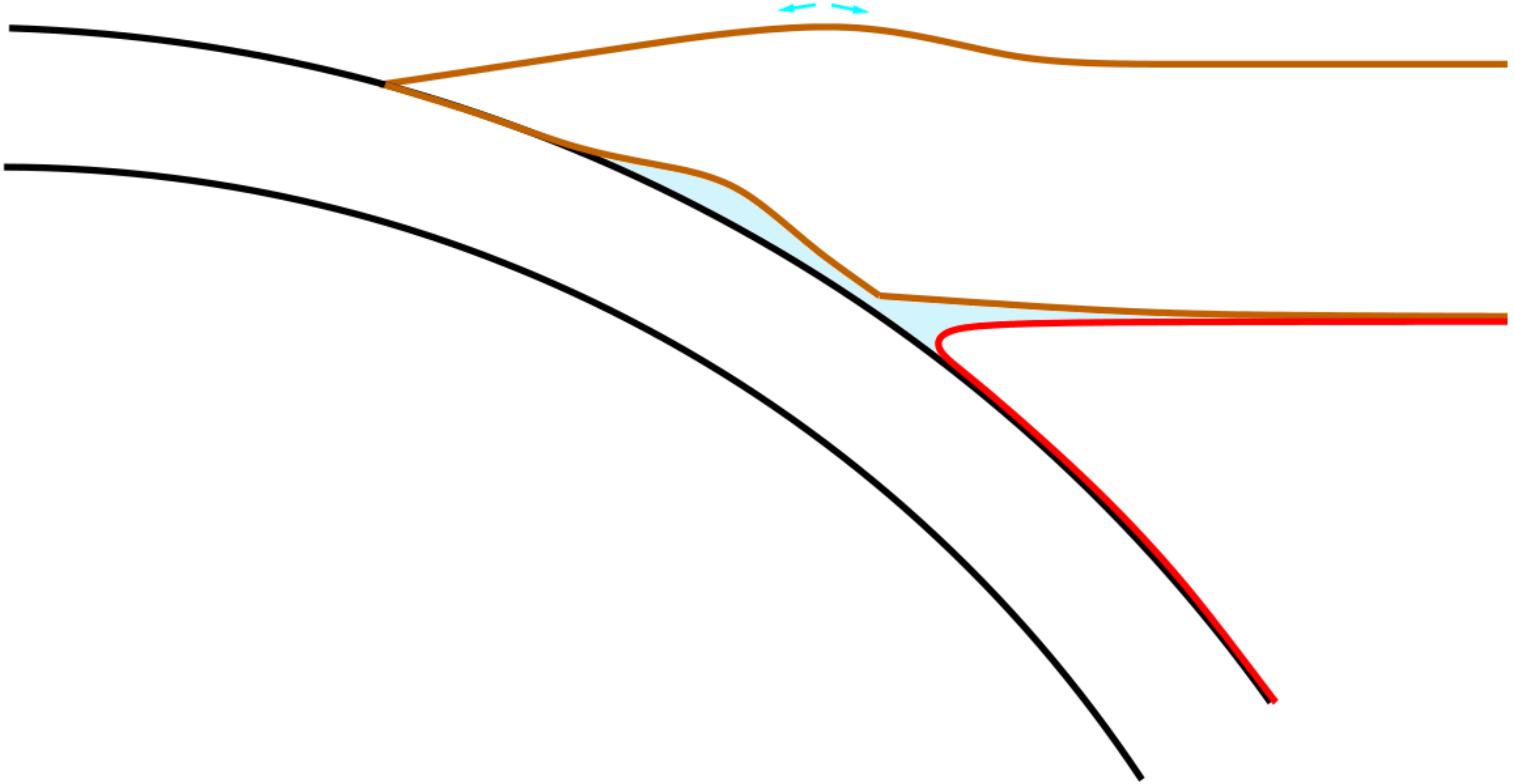}
		\thicklines
		\put(5,24){\color{black}\scalebox{1}{\rotatebox[origin=c]{169}{$\rightarrow$}}}
		\put(50,7){\color{black}\scalebox{1}{\rotatebox[origin=c]{145}{$\rightarrow$}}}
		\put(45,24){\color{red}\scalebox{2}{$\leftarrow$}}
		\put(51,20){\color{blue}\scalebox{2}{$\downarrow$}}
		\put(23,21){\color{black}\scalebox{2}{\rotatebox[origin=c]{70}{$\Big\updownarrow$}}}
		\put(23,18){\color{black}\scalebox{1}{\rotatebox[origin=c]{-20}{\myfont ?locked?}}}
		\put(85,17){\textbf{\myfont T$\bm{_3}$}}
		\end{overpic}}\\[5mm]
	\subfloat{\label{fig:T1}
		\begin{overpic}[width=0.75\textwidth,trim=20mm 70mm 30mm 0,clip=true]{EBC_Stage1_Ed.pdf}
		\thicklines
		\put(5,20){\color{black}\scalebox{2}{\rotatebox[origin=c]{-11}{$\rightarrow$}}}
		\put(50,3){\color{black}\scalebox{2}{\rotatebox[origin=c]{-35}{$\rightarrow$}}}
		\put(55,20){\color{red}\scalebox{1}{$\rightarrow$}}
		\put(53,24){\color{blue}\scalebox{1}{$\uparrow$}}
		\put(23,21){\color{black}\scalebox{2}{\rotatebox[origin=c]{70}{$\Big\updownarrow$}}}
		\put(23,19){\color{black}\scalebox{1}{\rotatebox[origin=c]{-20}{\myfont locked}}}
		\put(85,17){\textbf{\myfont T$\bm{_1}$}}
		\end{overpic}}\hfill
	\subfloat{\label{fig:T4}
		\begin{overpic}[width=0.75\textwidth,trim=20mm 70mm 30mm 0,clip=true]{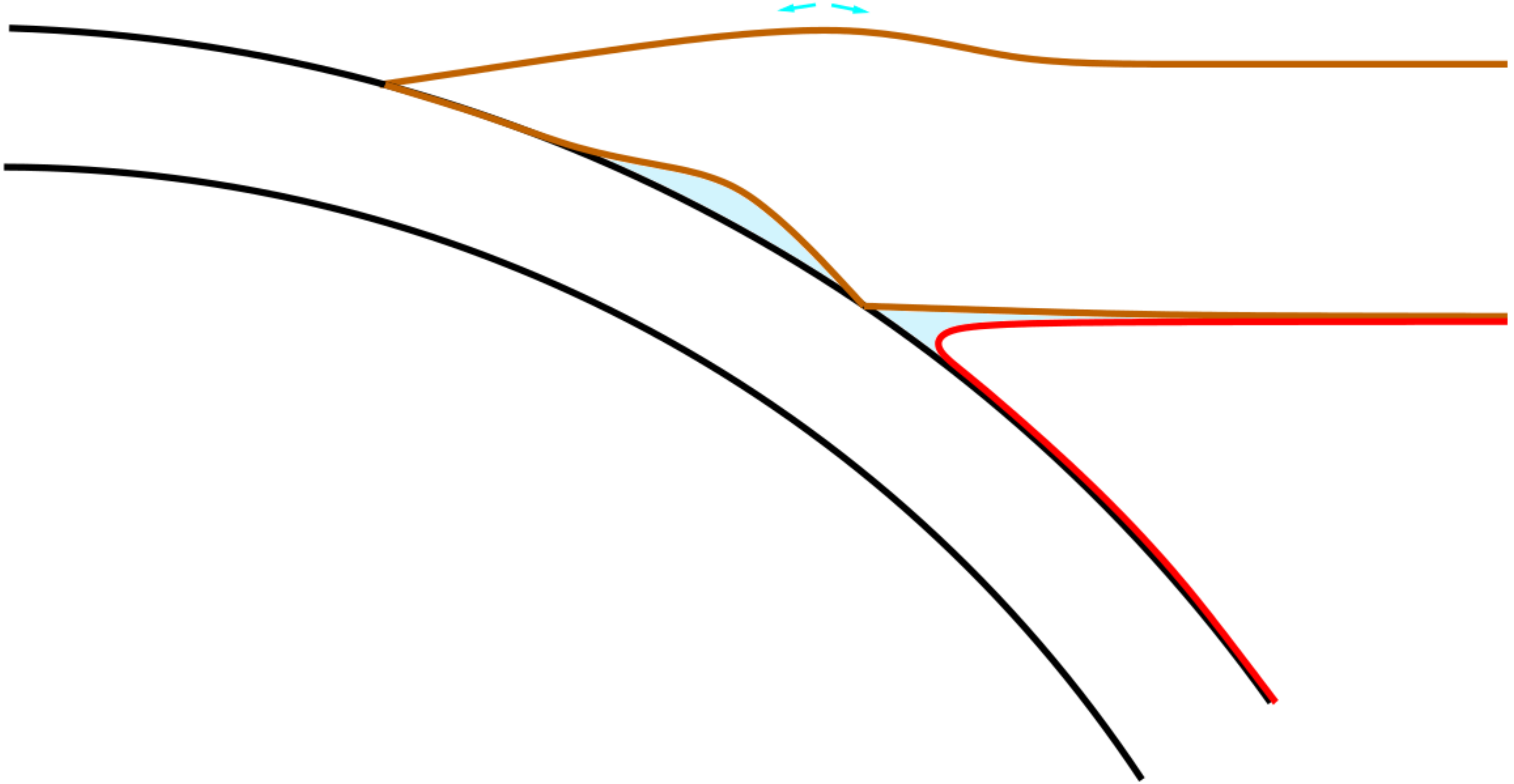}
		\thicklines
		\put(5,24){\color{black}\scalebox{1}{\rotatebox[origin=c]{169}{$\rightarrow$}}}
		\put(50,7){\color{black}\scalebox{1}{\rotatebox[origin=c]{145}{$\rightarrow$}}}
		\put(45,24){\color{red}\scalebox{1}{$\leftarrow$}}
		\put(51,20){\color{blue}\scalebox{1}{$\downarrow$}}
		\put(23,21){\color{black}\scalebox{2}{\rotatebox[origin=c]{70}{$\Big\updownarrow$}}}
		\put(23,18){\color{black}\scalebox{1}{\rotatebox[origin=c]{-20}{\myfont locked?}}}
		\put(59,8){\includegraphics[width=1mm]{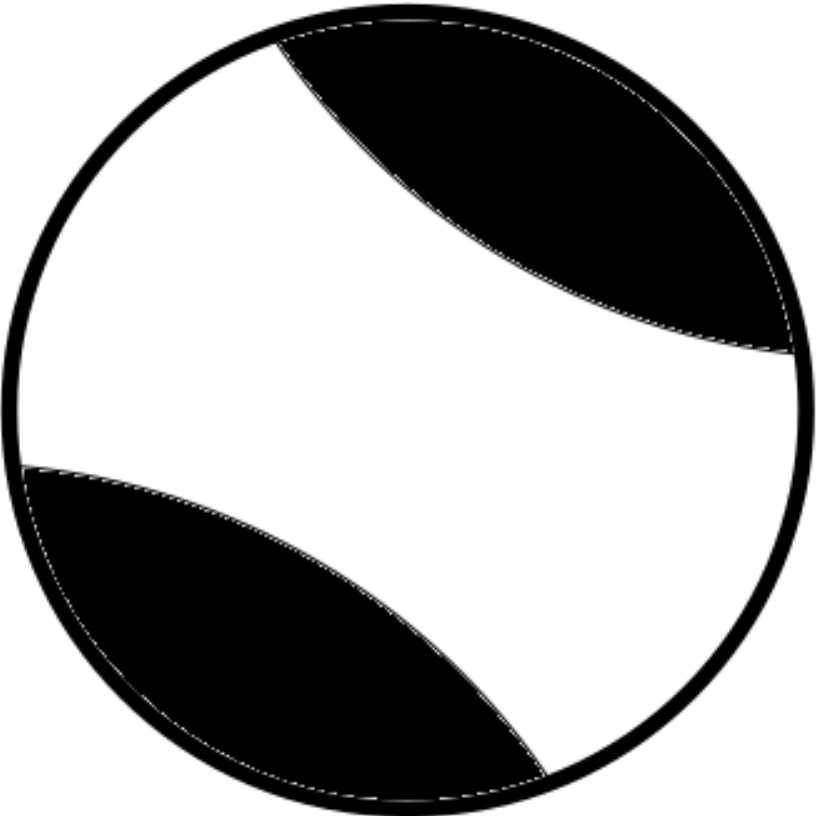}}
		\put(85,17){\textbf{\myfont T$\bm{_4}$}}
		\end{overpic}}\\[5mm]
	\subfloat{\label{fig:T2}
		\begin{overpic}[width=0.75\textwidth,trim=20mm 70mm 30mm 0,clip=true]{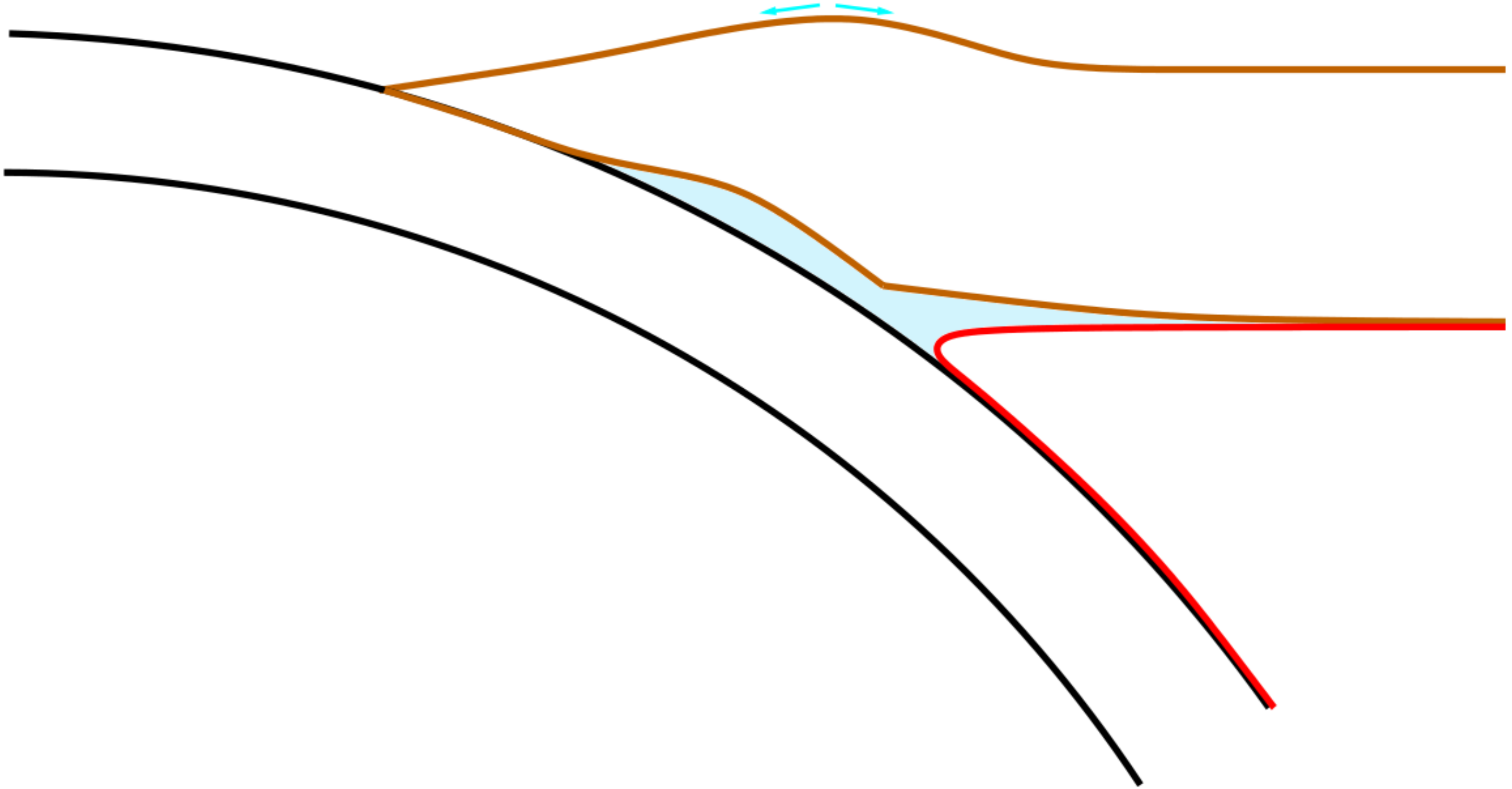}
		\thicklines
		\put(10,22){\color{black}{\circle*{1}}}
		\put(50,7){\color{black}{\circle*{1}}}
		\put(55,23){\color{red}\circle*{1}}
		\put(54,24){\color{blue}\circle*{1}}
		\put(23,20){\color{black}\scalebox{2}{\rotatebox[origin=c]{70}{$\Big\updownarrow$}}}
		\put(23,18){\color{black}\scalebox{1}{\rotatebox[origin=c]{-20}{\myfont locked?}}}
		\put(85,17){\textbf{\myfont T$\bm{_2}$}}
		\end{overpic}}\hfill
	\subfloat{\label{fig:T5}
		\begin{overpic}[width=0.75\textwidth,trim=20mm 70mm 30mm 0,clip=true]{EBC_Stage0_Ed.pdf}
		\thicklines
		\put(10,22){\color{black}{\circle*{1}}}
		\put(50,7){\color{black}{\circle*{1}}}
		\put(55,23){\color{red}\circle*{1}}
		\put(54,24){\color{blue}\circle*{1}}
		\put(23,21){\color{black}\scalebox{2}{\rotatebox[origin=c]{70}{$\Big\updownarrow$}}}
		\put(23,18){\color{black}\scalebox{1}{\rotatebox[origin=c]{-20}{\myfont locked?}}}
		\put(53,12){\includegraphics[width=1mm]{Eq_CLVD_Side.pdf}}
		\put(56,10){\includegraphics[width=1mm]{Eq_CLVD_Side.pdf}}
		\put(59,8){\includegraphics[width=1mm]{Eq_CLVD_Side.pdf}}
		\put(85,17){\textbf{\myfont T$\bm{_5}$}}
		\end{overpic}}
	\vspace{-2mm}
	\caption{\textbf{Schematic illustration of the different phases of the Episodic Buckling and Collapse model of the subduction process and the structural changes therein.} The subducting oceanic crust is outlined by black lines and the black arrows represent the direction and magnitude of the slab velocity. The overriding continental crust is represented by the solid brown lines. Red and blue arrows represent the magnitudes of the instantaneous horizontal and vertical velocities, respectively, of a point in the continental crust wedge. Dots in Phases \textbf{T}$\bm{_2}$, and \textbf{T}$\bm{_4}$ represent vectors of magnitude zero. The tilt magnitude and direction are denoted by the arrows in cyan. The side-view of CLVD focal spheres are shown along the plate interface. Temporal motion of locations A through D on the continental crust surface are analyzed in Figure~\ref{fig:GPSandHodo} below.
}
	\label{fig:Buckling}
\end{figure}
\end{landscape}

\textbf{Up-dip Tremor Migration} -- 
In addition, as supporting frictional forces are overcome, the lower portion of the continental crust wedge strikes the subducting slab first (phase \textbf{T}$\bm{_4}$, Figure~\ref{fig:Buckling}), followed by a progressive collapse of the continental crust along the up-dip (and radial) direction (phase \textbf{T}$\bm{_5}$, Figures~\ref{fig:Buckling} and \ref{fig:MagmaFlow_T5}) -- interpreted as up-dip and radial tremor migration in several studies \cite{Wech2009,Obara2011b}.

\textbf{Along-strike Tremor Migration} -- 
The locked zone prevents the fluid pocket from moving further up-dip and therefore the fluid pocket migrates parallel to the margin of the locked zone as depicted in Figure~\ref{fig:MagmaFlow_T6}.
In phase \textbf{T}$\bm{_6}$, we believe that the trapped fluids move predominantly along-strike resulting in the observed along-strike tremor migration patterns \cite{Wech2009,Obara2011b,Obara2012}.
Fluids are pushed along-strike until they are lost to the overlying permeable crust and/or are pushed down along the plate interface.
Because of the progressive loss in fluid-pressure in the latter stages, the rate of along-strike collapse is expected to be lower than the initial up-dip collapse rate -- which explains the slower along-strike tremor migration with respect to up-dip migration \cite{Wech2009,Houston2011,Obara2011b,Obara2012}.
This model also explains the bimodal distribution of tremors in the Nankai subduction zone \cite{Obara2012} with tremors from the along-strike migration concentrated on the up-dip side while tremors from up-dip migration are distributed over the entire tremor zone.

Some studies \cite{Houston2011,Obara2012} also report rapid reverse tremor migration where tremors migrate in the opposite direction of along-strike migration at much faster speeds.
We postulate that rapid tremor reversal happens when a migrating high-pressure fluid pocket encounters a permeable zone such as a fault or fracture zone, or a magma vent or dike.
As fluid escapes through these fissures, the leading edge of the fluid pocket collapses rapidly.
This collapse is in the direction opposite to the migrating fluid front and occurs at a much faster rate given the loss of pore pressure in the fluid pocket.

\emph{Note that the fluid chamber does not fully collapse within each cycle, instead there is a partial collapse. However, with each passing cycle we expect a net increase in the fluid chamber size from one EBC cycle to the next. Only when the frictional forces in the locked zone are overcome during a megathrust earthquake, does the fluid chamber completely collapse. This fluid-filled chamber is responsible for attenuation of high-frequency portion on the foreshocks as observed by \citeA{PinaValdes2018a} and \citeA{PinaValdes2018b}.}

\textbf{Further Evidence of Fluid Flow} -- 
The periodic changes in seismicity rates and attenuation and their correspondence with accelerated slow slip, as reported by \citeA{Nakajima2018}, corroborates the above model of fluid flow in and out of the fluid chamber.
The `breathing' mechanism of magmatic fluid flow driven by periodic plate deformation in subduction zones might be the dominant mechanism (and not buoyancy) of magma transport from the upper mantle to the crust and might even be responsible for the creation of the Aleutian Volcanic Arc in Alaska and its volcanism as evident from the focusing of partial melt under the arc.

\begin{landscape}
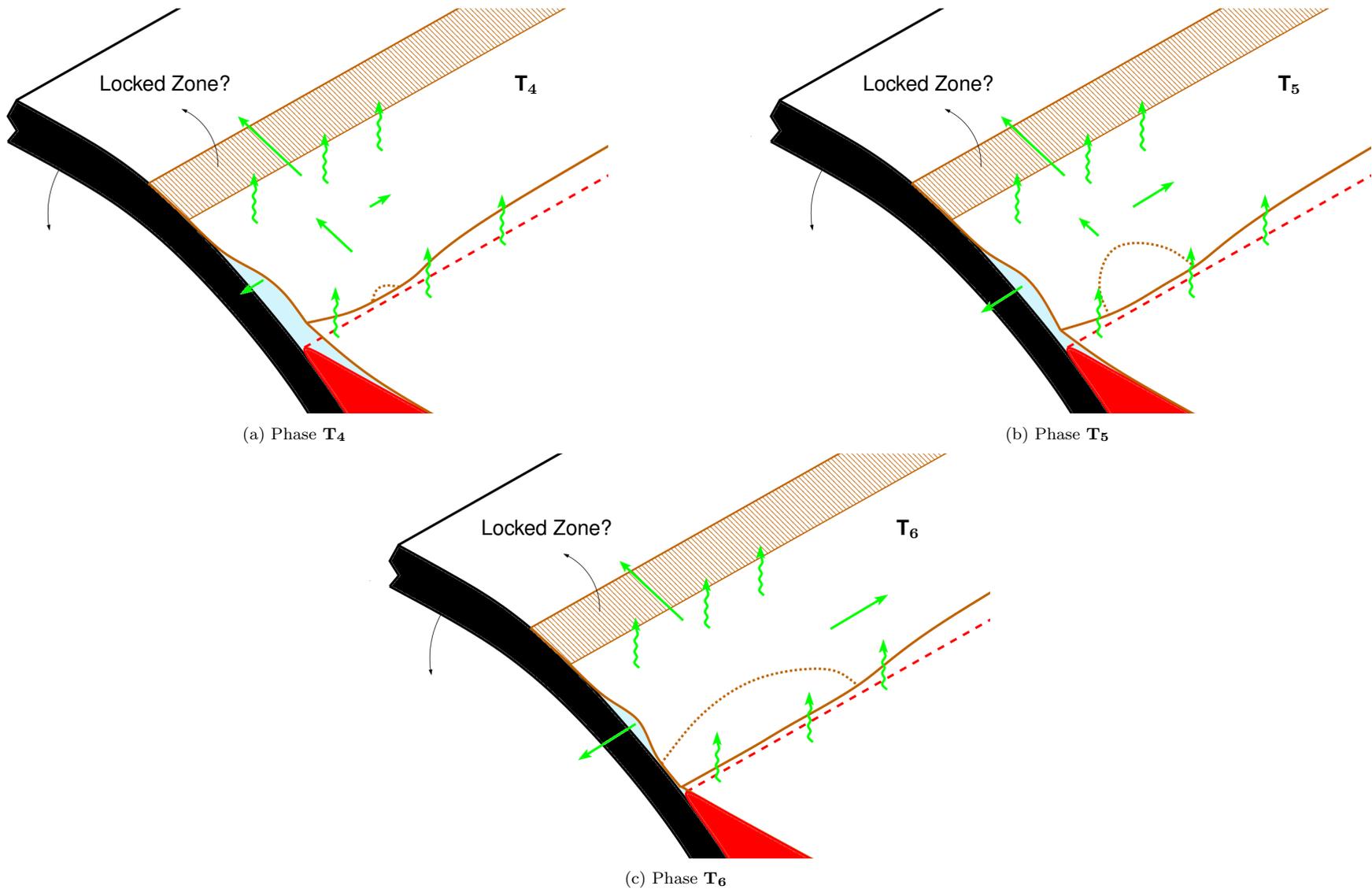
\begin{figure}
	\center
	\subfloat[Phase \textbf{T}$\bm{_4}$]{\label{fig:MagmaFlow_T4}
		\begin{overpic}[width=0.75\textwidth,trim=0mm 60mm 80mm 90mm,clip=true]{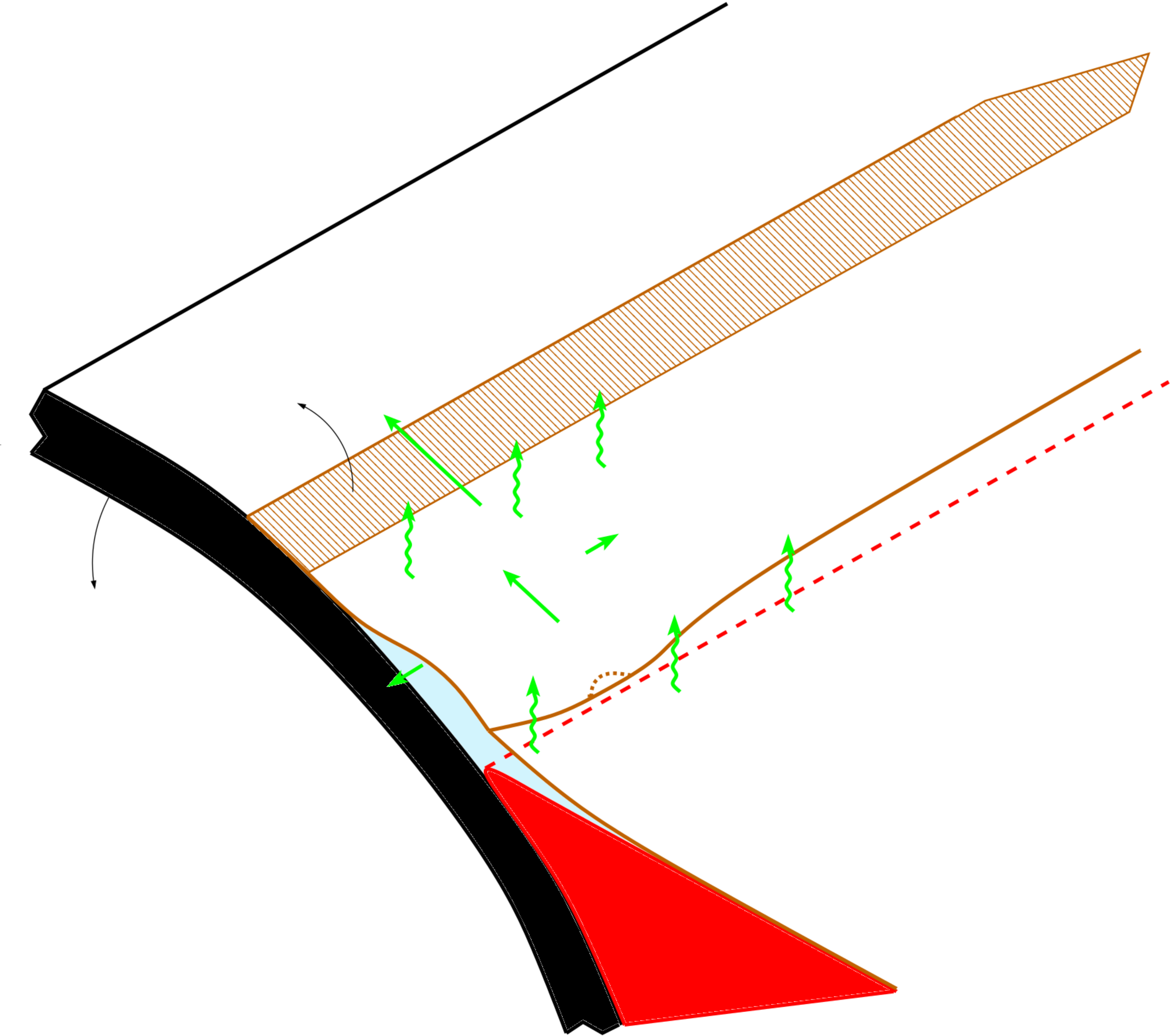}
		\put(18,52){\myfont Locked Zone?}
		\put(85,52){\textbf{\myfont T$\bm{_4}$}}
		\end{overpic}}\hfill
	\subfloat[Phase \textbf{T}$\bm{_5}$]{\label{fig:MagmaFlow_T5}
		\begin{overpic}[width=0.75\textwidth,trim=0mm 60mm 80mm 90mm,clip=true]{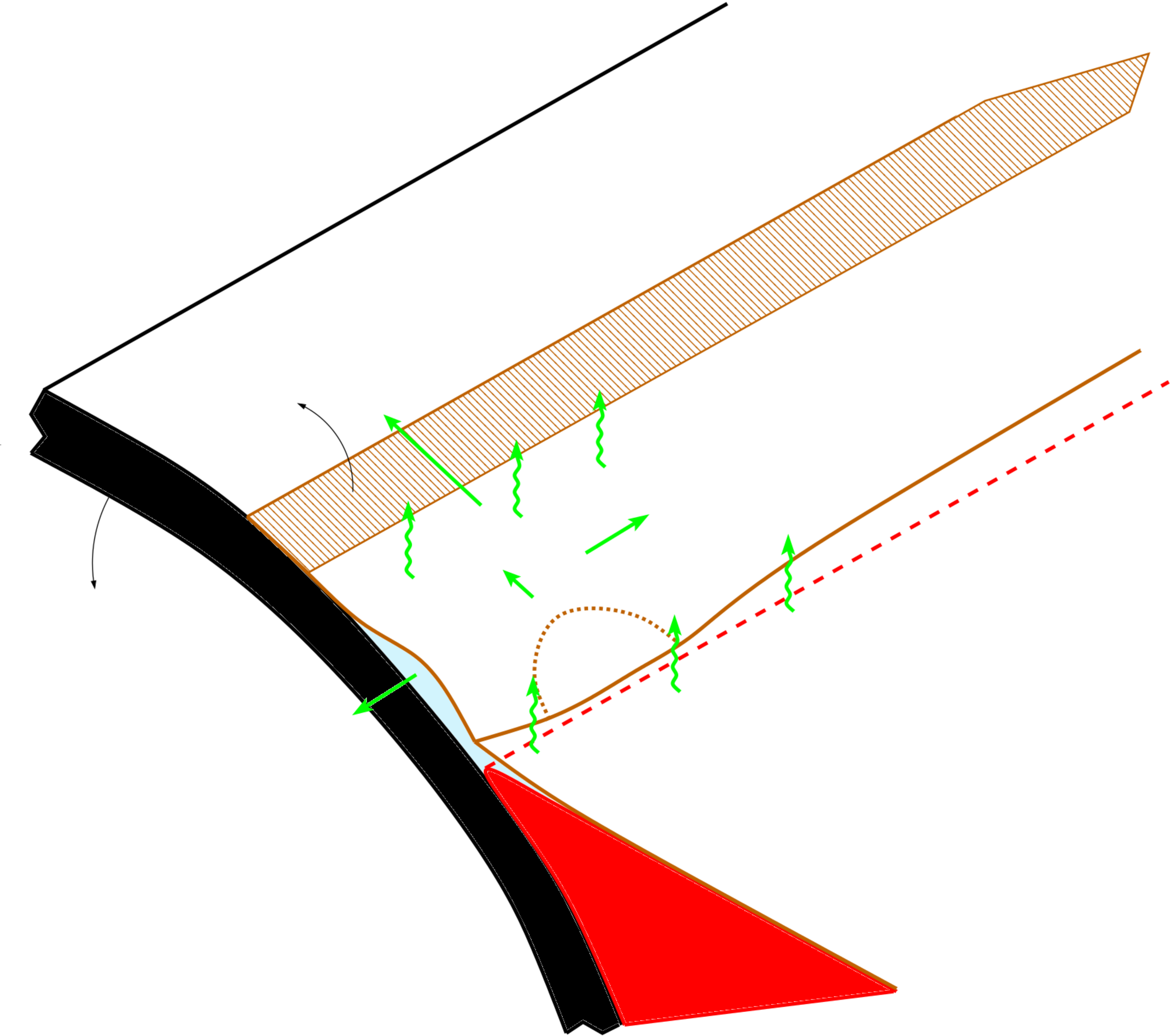}
		\put(18,52){\myfont Locked Zone?}
		\put(85,52){\textbf{\myfont T$\bm{_5}$}}
		\end{overpic}}\\[1mm]
	\subfloat[Phase \textbf{T}$\bm{_6}$]{\label{fig:MagmaFlow_T6}
		\begin{overpic}[width=0.75\textwidth,trim=0mm 60mm 80mm 90mm,clip=true]{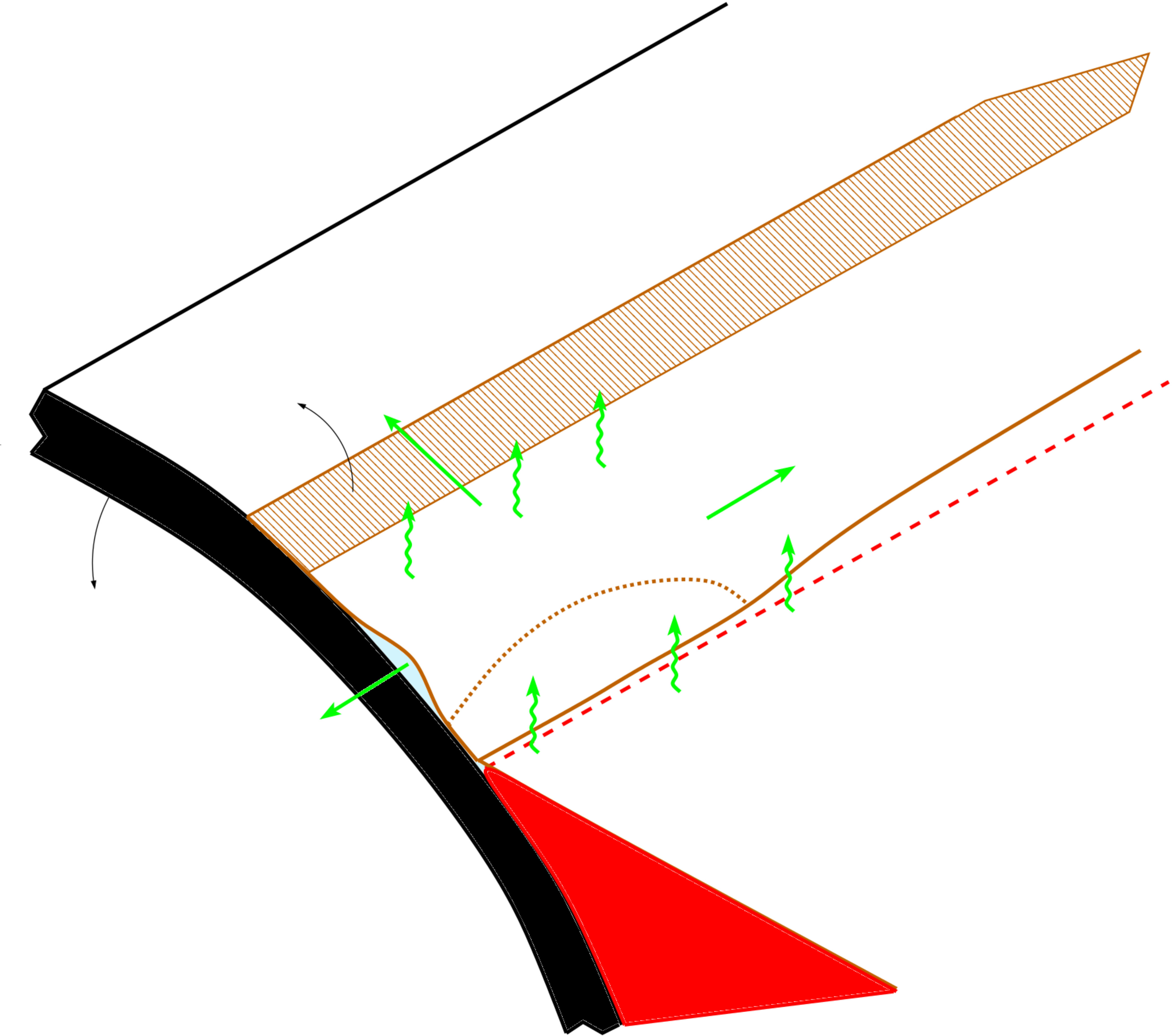}
		\put(18,52){\myfont Locked Zone?}
		\put(85,52){\textbf{\myfont T$\bm{_6}$}}
		\end{overpic}}
	\vspace{-2mm}
	\caption{\textbf{Three-dimensional schematic illustration of magmatic fluid flow during the latter stages of each Episodic Buckling and Collapse cycle.} Only the basal surface of the continental crust is shown for clarity. The dashed brown line corresponds to the leading edge of the advancing tremor front (edge of the contact between the continental crust and subducting slab). Green arrows represent the flow direction of magmatic fluids. The sinuous green lines fluid flow into the overlying crust.}
	\label{fig:MagmaFlow3D}
\end{figure}
\end{landscape}

\section{Megathrust Cycle Made up of Multiple EBC Cycles}\label{MegathrustEBC}
All existing megathrust models assume the continental crust to be in contact with the subducting slab and the mantle at all locations.
On the other hand, according to the Episodic Buckling and Collapse model, at the start of the megathrust buckling cycle, the continental crust is in direct contact with the subducting slab at all depths.

However, with each short-term buckling cycle, there is a net positive accumulation of vertical strain within the continental crust, resulting in progressive vertical detachment of the crust and slab as depicted below.
Although majority of the strain in the overriding plate is released when it collapses, a small portion of the strain is retained in every EBC cycle.
Over hundreds of short-term buckling and collapse cycles, the small retained strains add up and this strain energy is stored in the overriding continental plate.
During GPS processing, one usually performs a detrending step, which eliminates the signature corresponding to this stored strain energy.

\begin{landscape}
\begin{figure}
	\center
	\includegraphics[width=1.5\textwidth]{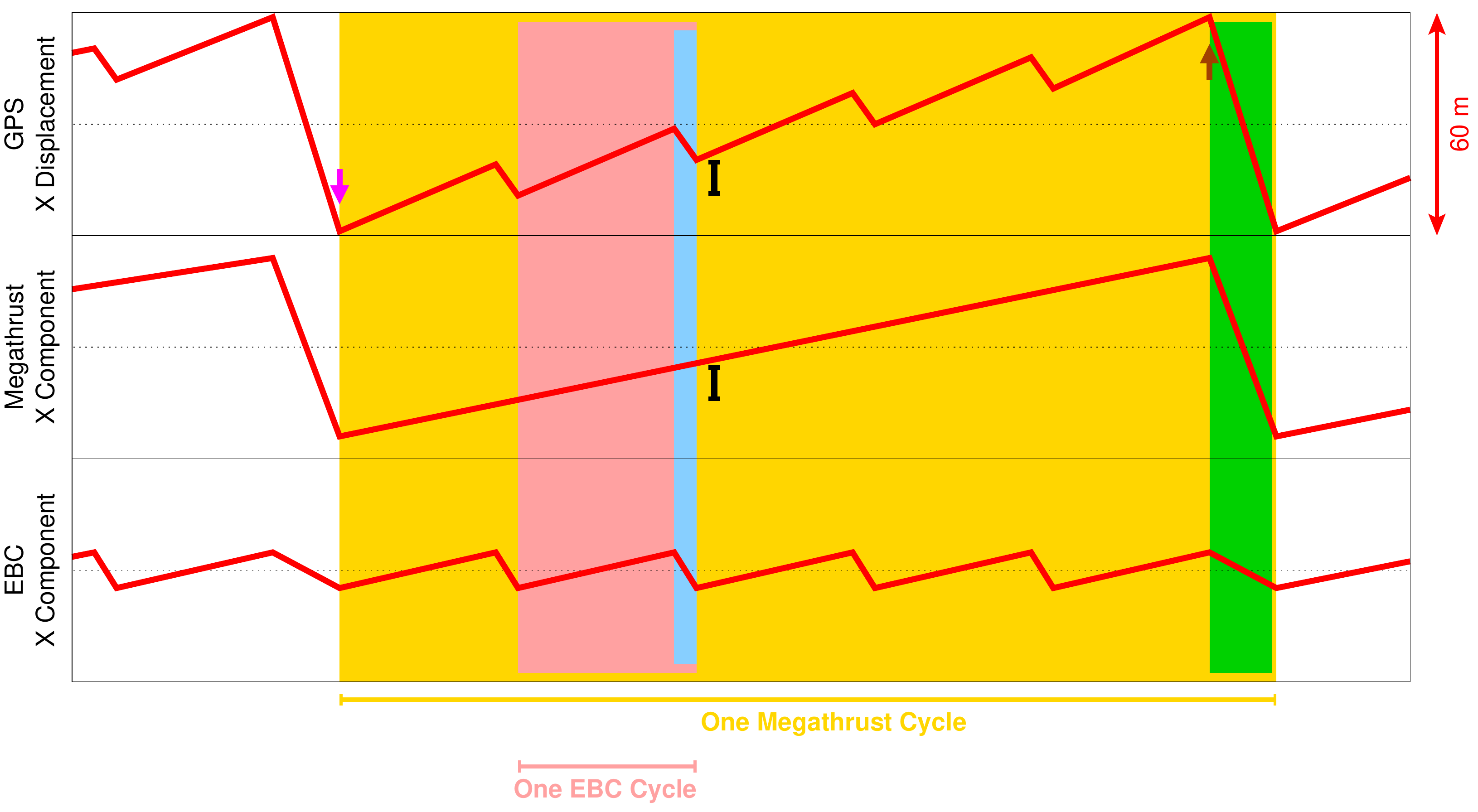}
	\caption{\textbf{Schematic of horizontal surface displacement of location A (near trench) observed on GPS (top plot), and decomposed into the megathrust component (middle plot) and the EBC component (bottom plot).} The yellow time period denotes a single megathrust cycle, with the green period corresponding to the megathrust afterslip. The pink time period demarcates one of the multiple EBC cycles, with the blue corresponding to the collapse phase of the EBC cycle. Each megathrust cycle is composed of several EBC cycles. For illustration purposes, we show only 5 EBC cycles constituting each megathrust cycle; in reality, each megathrust cycle constitutes hundreds of EBC cycles. The maximum horizontal displacement (often referred to as megathrust afterslip) for each megathrust cycle is approximately 60~m. During routine GPS processing, a detrending step is carried out which removes the imprint of the megathrust cycle. The black line segment represents the net horizontal strain accumulated during each EBC cycle (and removed with detrending) at location A. The brown arrow corresponds to the time of the megathrust earthquake, when the seismogenic zone gets unlocked and the overriding plate begins its complete collapse. The magenta arrow represents the beginning of the megathrust cycle when the overriding plate has completely collapsed on the subducting slab and is in full contact with it all along the dip line.}
	\label{fig:EqCyclesX}
\end{figure}
\end{landscape}

\begin{landscape}
\begin{figure}
	\center
	\includegraphics[width=1.5\textwidth]{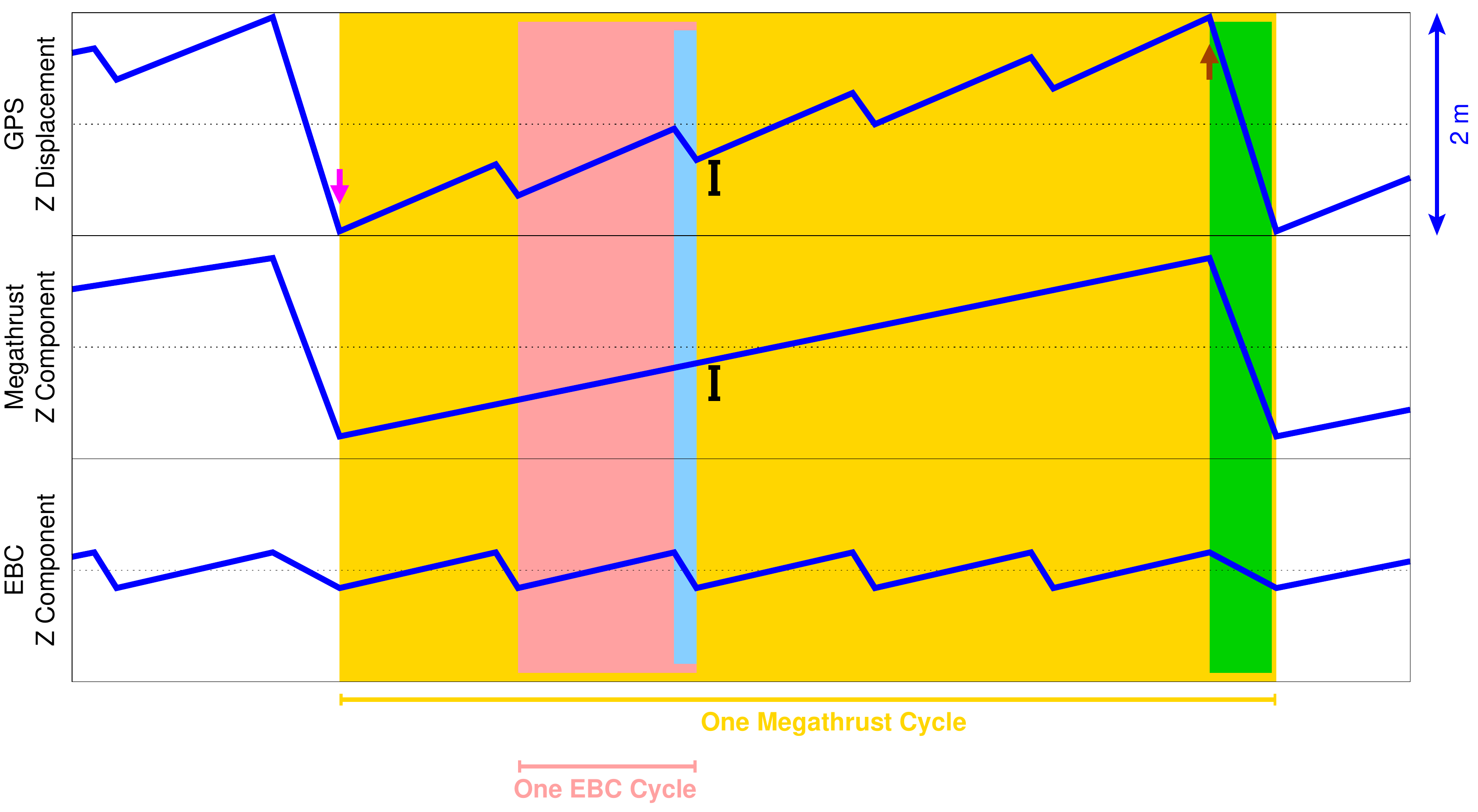}
	\caption{\textbf{Schematic of vertical surface displacement of location C (approximately above tremor zone) observed on GPS (top plot), and decomposed into the megathrust component (middle plot) and the EBC component (bottom plot).} As in figure~\ref{fig:EqCyclesX}, the yellow time period denotes a single megathrust cycle, with the green period corresponding to the megathrust afterslip. The pink time period demarcates one of the multiple EBC cycles, with the blue corresponding to the collapse phase of the EBC cycle. Each megathrust cycle is composed of several EBC cycles. For illustration purposes, we show only 5 EBC cycles constituting each megathrust cycle; in reality, each megathrust cycle constitutes hundreds of EBC cycles. The maximum vertical displacement, as seen in the work of \protect\citeA{Sherrod2001,Leonard2004,Hamilton2005a,Hamilton2005b,Shennan2006}, for each megathrust cycle is approximately 2~m. During routine GPS processing, a detrending step is carried out which removes the imprint of the megathrust cycle. The black line segment represents the net vertical strain accumulated at location C during each EBC cycle (and removed with detrending). Again, the brown arrow corresponds to the time of the megathrust earthquake, when the seismogenic zone gets unlocked and the overriding plate begins its complete collapse. The magenta arrow represents the beginning of the megathrust cycle when the overriding plate has completely collapsed on the subducting slab and is in full contact with it all along the dip line.}
	\label{fig:EqCyclesZ}
\end{figure}
\end{landscape}

A critical state is attained where the forces exerted by the stored elastic energy (due to compression) and gravitational potential energy (stored in the uplifted continental crust) exceed the frictional forces in the seismogenic zone (locked zone).
This state of deformation exhibits the maximal horizontal (at location A) and vertical (at location C)  displacements of the overriding plate (Figures~\ref{fig:EqCyclesX} and \ref{fig:EqCyclesZ}, respectively).
As the frictional forces are exceeded, the stored energy is released in the form of a megathrust earthquake (end of megathrust cycle in Figures~\ref{fig:EqCyclesX} and \ref{fig:EqCyclesZ}).

We hypothesize that the short-term buckling and collapse cycles described above are sequences that make up each long-term megathrust cycle. Therefore, each megathrust cycle can be considered to be one centuries-long buckling and collapse cycle which in turn is made up of numerous short-term cycles.
Evidence of these inter-seismic vertical crustal deformations corresponding to megathrust earthquakes is found in long-term geologic records \cite{Dragert1994,Sherrod2001,Leonard2004,Hamilton2005a,Hamilton2005b,Shennan2006} and may be interpreted as large time-scale versions of the buckling process that take centuries to develop.
The rapid subsidence, observed by \citeA{Sherrod2001,Leonard2004,Hamilton2005a,Hamilton2005b,Shennan2006} on geologic records, occurs during the express subsidence of the overriding continental plate.
Most researchers report a maximum vertical subsidence of 2 to 3~m (Figure~\ref{fig:EqCyclesX}).

As the continental plate completely collapses after a megathrust earthquake, the horizontal component of the GPS shows large seaward displacements (in particular the locations close to the trench, Figure~\ref{fig:EqCyclesX}).
The large aseismic afterslip, following megathrust earthquakes and observed in multiple studies \cite{Gomberg2012,Rolandone2018}, is simply the horizontal projection of the seaward surface displacement of the overriding continental plate while it completely collapses.
Several researchers report displacements as large as 60~m.
Also, because the overriding plate gradually collapses while pushing fluids out (instead of sliding on the oceanic slab), there is no seismic energy released -- it is predominantly aseismic.
The magmatic fluids are most likely pushed out along-strike and to the trench along the ruptured plate boundary as evidenced by the significant increase in mantle helium in the seawater and reported by \citeA{Sano2014}.

As suspected by several geoscientists, the periodic release of stored energy in subduction zones in the form of fluid flow and seismic events, during each Episodic Buckling and Collapse cycle, indeed prevents megathrust earthquakes from occurring more frequently.
A back-of-the-envelope calculation shows that if not for the episodic energy release, the Cascadia region would be experiencing one megathrust earthquake every 54~years.

Therefore, we believe that the key to forecasting megathrust earthquakes in a cost-effective fashion is to monitor long-term trends (in the order of decades and centuries) in ground deformation through multi-component GPS and tiltmeter recordings.

\acknowledgments
Partial funding for is provided by the National Science Foundation under the EAGER program (NSF Grant 1933169).
In Cascadia, GPS time series are provided by the Pacific Northwest Geodetic Array, Central Washington University (https://www.geodesy.cwu.edu/).
For Alaska, GPS time series data were downloaded from USGS (https://earthquake.usgs.gov/monitoring/gps).
We thank Manika Prasad for discussions.

\bibliography{/home/jbehura/Work/MyTex/References}

\end{document}